\documentclass[preprints,article,accept,moreauthors,pdftex]{Definitions/mdpi}
\firstpage{1} 
\pubvolume{1}
\issuenum{1}
\articlenumber{0}
\pubyear{2022}
\copyrightyear{2022}
\datereceived{} 
\dateaccepted{} 
\datepublished{} 
\hreflink{https://doi.org/} % If needed use \linebreak
\pdfoutput=1
\Title{Acoustic, phononic, Brillouin light scattering and Faraday wave
based frequency combs: physical foundations and applications}

\TitleCitation{Acoustic, phononic, Brillouin light scattering and Faraday wave
based frequency combs}

\newcommand{\M}{{\cal M}}
\newcommand{\W}{{\cal W}}
\newcommand{\R}{{\cal R}}
\newcommand{\Q}{{\cal Q}}
\newcommand{\K}{{\cal K}}

\newcommand{\cO}{{\cal O}}

\usepackage{textcomp}
\usepackage{bm}
\usepackage{comment}

\Author{Ivan S.~Maksymov $^{1}$*\orcidA{}, Bui Quoc Huy Nguyen $^{1}$\orcidB{},
Andrey Pototsky $^{2}$\orcidC{} and Sergey A.~Suslov $^{2}$\orcidB{}}

\AuthorNames{Ivan S.~Maksymov, Bui Quoc Huy Nguyen,
Andrey Pototsky and Sergey A.~Suslov}

\AuthorCitation{Maksymov, I.~S.; Nguyen, Bui Quoc Huy Nguyen;
  Pototsky, A.; Suslov, S.~A.} 
\address{%
$^{1}$ \quad Optical Sciences Centre, Swinburne University of 
Technology, Hawthorn, Victoria 3122, Australia\\ 
$^{2}$ \quad Department of Mathematics, Swinburne University of
  Technology, Hawthorn, Victoria 3122, Australia}

\corres{Correspondence: imaksymov@swin.edu.au ; Tel.: +61-3-3921-4805}

\abstract{Frequency combs (FCs)---spectra containing equidistant
coherent peaks---have enabled researchers and engineers to
measure the frequencies of complex signals with high precision
thereby revolutionising the areas of sensing, metrology and
communications and also benefiting the fundamental science. 
Although mostly optical FCs have found widespread applications
thus far, in general FCs can be generated using waves other
than light. Here, we review and summarise recent achievements in
the emergent field of acoustic frequency combs (AFCs) including
phononic FCs and relevant acousto-optical, Brillouin light
scattering and Faraday wave-based techniques that have
enabled the development of phonon lasers, quantum computers and
advanced vibration sensors. In particular, our discussion is
centred around potential applications of AFCs in precision
measurements in various physical, chemical and biological
systems in conditions, where using light, and hence optical
FCs, faces technical and fundamental limitations, which is,
for example, the case in underwater distance measurements and
biomedical imaging applications. This review article 
will also be of interest to readers seeking a discussion 
of specific theoretical aspects of different classes of AFCs.
To that end, we support the mainstream discussion by the results
of our original analysis and numerical simulations that can be used to
design the spectra of AFCs generated using oscillations of gas bubbles
in liquids, vibrations of liquid drops and plasmonic enhancement of
Brillouin light scattering in metal nanostructures. We also discuss
the application of non-toxic room-temperature liquid-metal alloys in
the field of AFC generation.}


\begin{document}
%%%%%%%%%%%%%%%%%%%%%%%%%%%%%%%%%%%%%%%%%%
\section{Introduction and motivation\label{sec:1}}
Precision measurement underpins modern technologies that are
critical for timing and communication as well as for fundamental
science such as astrophysics. However, electronic and optical
devices used in measurement systems generate interference (noise)
that complicates or even prevents reading a pure signal. Thus, one
of the main goals of precision measurement is to reduce the noise
level by improving the signal-to-noise ratio and increasing the
sensitivity of sensors and signal detectors.

Advances in optical technologies are essential for achieving these
goals. Indeed, the unique physical properties of light and recent
progress in the development of novel sources of light and synthesised 
optical materials enabling light manipulation at nanoscale open
unprecedented opportunities for reducing noise levels and increasing
the measurement accuracy. For example, optical frequency combs (OFCs)
(Sec.~\ref{sec:2}) have enabled scientists and engineers to measure
and control light waves as if they were radio waves. Using OFCs,
established technologies that employ radio and microwave
frequencies---clocks, computers and telecommunications systems---can
be seamlessly connected to devices that use optical waves with frequencies
approximately 10,000 times higher than those of radio and microwaves
\cite{Pic19, Wei19, For19}. 

However, while optical technologies are an invaluable tool that
researchers use to explore new horizons, they have a number of
drawbacks that originate from fundamental physical limits. This is, for
example, the case in underwater communication \cite{Uri83, Wu19}
and of some medical imaging and sensing modalities used deeply inside a
living human body \cite{Wan10, Mak16}, where the intensity of light is
dramatically attenuated due to scattering and optical absorption
in liquids and bodily fluids and tissues. This situation has motivated
pioneering studies of alternative approaches that use waves other than
light to enable sensing and precision measurement in specific but
critical areas of health studies and deep sea exploration. Given that optical 
waves share many fundamental physical properties with the waves of other
nature---most notably with sound waves \cite{Mak19}---it has been
suggested that certain optical precision measurement and sensing
technologies, including OFCs, could be implemented using acoustic
waves and vibrations.

In this review article, we critically review results of recent
studies, where novel acoustic frequency combs (AFC)---non-optical 
counterparts of OFCs---have been introduced and a number of their
potential applications suggested. We pursue several goals in the
present review. Firstly, since the field of AFCs is relatively new,
the terminology used in it is still not uniform and can vary from one
paper to another (for example, although throughout the mainstream
discussion we will use the term AFC, some authors prefer calling
them phononic frequency combs \cite{Gan17}). Subsequently, one of
our aims is to provide a taxonomy of AFCs and relevant concepts of
non-optical frequency combs (FCs). Secondly, even though the seminal
works on distinct non-optical FC technologies appeared almost
simultaneously, they have resulted from rather isolated research
efforts. While some of the AFC investigators have already consolidated
their activities, which is, for example, the case in optomechanics and
Brillouin light scattering communities \cite{Lae15, montes, Egg19,
  Lau19} working on integrated photonic circuits \cite{Kan11, She18},
in general there is still no coherent research framework for the
future development of AFCs. Therefore, the current review intends to
promote collaboration between different research groups. Given this,
the mainstream discussion of this article will specifically focus on the
results that have received a limited attention thus far but that, in
our opinion, hold the promise to find their own application niche and
influence further developments in the adjacent areas. Finally, to make
the article accessible to non-specialists, the discussion of each
specific AFC technique is accompanied by an overview of relevant 
physical phenomena with suggestions of further reading for researchers
interested in more detail.  
     
\section{Optical frequency combs\label{sec:2}}
An optical frequency comb (OFC) is a spectrum consisting of a series
of discrete, equally spaced elements with a well-defined phase
relationship between them. The foundations of this breakthrough
technology were laid in the works led by the co-recipients of the 2005
Nobel Prize in Physics John Hall and Theodor H{\"a}nsch. Their
contributions to the field of precision spectroscopy enabled measuring
the light frequency with an unprecedented accuracy \cite{Hal00, Han06,
  Hal06}. They and their collaborators demonstrated that a stable laser
emitting light with a spectrum containing very fine colour (frequency)
lines can be used in combination with an FC technique to measure the
frequency of light very accurately. They also suggested that employing
an FC technique enables measuring both time and distance more
accurately than using any other approach.

In practice, the light frequency $f$ that needs to be to determined
may be too high to be measured directly. Subsequently, an indirect
measurement technique was proposed, where one compares the unknown
frequency to an optical ruler---an OFC \cite{Pic19, Wei19, For19}. The
comparison between two light frequencies is made using a well-known
beat technique that is based on measuring a frequency difference
between the known and investigated waves, which is sufficiently small
to be reliably measured using conventional methods.

\begin{figure}[t]
  \centering
  \includegraphics[width=12.0 cm]{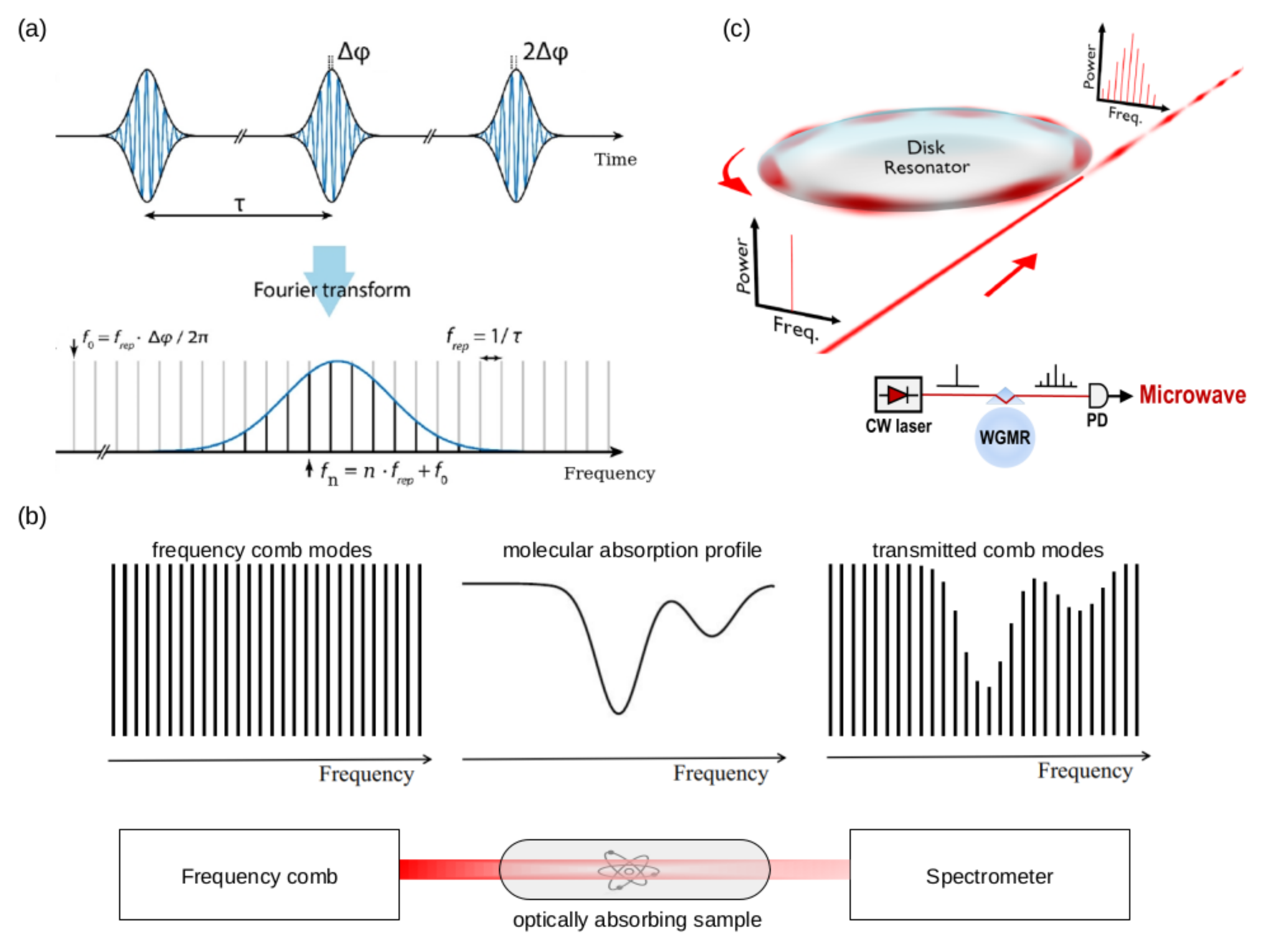} 
  \caption{\textbf{(a)}~Schematic of the OFC generation using a
    mode-locked laser. The top panel shows a train of optical pulses
    with a period $1/f_{rep}$. The bottom panel depicts the spectrum
    of narrow frequency peaks corresponding to the train of pulses in
    the time domain. The phase shift $\Delta\phi$ of the carrier wave
    with respect to the pulse envelope induces a translation
    $f_0=f_{rep}\Delta\phi/2\pi$ of spectral peaks from their harmonic
    frequencies $nf_{rep}$. Reproduced from \cite{Wei19} with permission from Elsevier.
    \textbf{(b)}~Sketch of an OFC-based spectroscopy technique.
    The OFC as a broadband light source interrogates an absorbing sample
    and a spectrometer analyses the transmission spectrum.  
    \textbf{(c)}~Artist's rendition of a Kerr effect-based OFC
    generation in a micro-photonic disc resonator. A continuous-wave
    (CW) input pump wave creates a pattern of optical
    whispering-gallery modes (WGMs) inside the disc resonator, thereby
    inducing a periodic light intensity modulation and thus producing
    an output signal with an OFC-like spectrum. The bottom panel shows
    the sketch of a simplified experimental setup for the generation
    of a Kerr OFC and its further processing using a photodetector
    (PD). Reproduced from \cite{Che16} published by De Gruyter Open 
    under the Creative Commons Attribution-NonCommercial-NoDerivs 3.0
    License.\label{Fig1}}
\end{figure}

An OFC is typically generated using a mode-locked laser system
\cite{Hal01}, where, in the time domain [Fig.~\ref{Fig1}(a)]
\cite{Pic19, Wei19}, a train of ultrashort optical pulses is emitted. The
period of the pulse envelope $1/f_{rep}=L/v_g$ corresponds 
to a round-trip time inside the laser cavity with a round-trip
length $L$ and the group speed of light $v_g$. Due to dispersion of
light inside the cavity, there is a phase shift $\Delta\phi$ between
the carrier and the pulse envelope signals. Hence, in the frequency
domain [Fig.~\ref{Fig1}(a)], the corresponding optical spectrum
consists of a discrete set of equidistant narrow peaks with
frequencies $f_n=nf_{rep}+f_0$, where $n$ is a large integer
indicating that the number of the peaks can be very high, $f_{rep}$ is
the repetition frequency of the pulse envelope and $f_0$ is the 
carrier-envelope offset frequency that is related to the phase shift
$\Delta\phi$ as $f_0=f_{rep}\Delta\phi/2\pi$ \cite{Pic19, Wei19}.

Significantly, soon after the introduction of the concept of OFC, this
technique found numerous applications beyond its originally intended
use. For example, it was established that OFCs could provide 
long-term calibration of essential astronomical equipment \cite{Wil12},
enable flexible control of ultrashort optical pulses \cite{Bal03}
and benefit generation of arbitrary radio-frequency waveforms and
optical communications \cite{Tor13}. Of particular importance is also
the application of OFCs is in the field of spectroscopy \cite{Ye05,
  Mad13}, where, in an idealised OFC-based system
[Fig.~\ref{Fig1}(b)], an OFC both optically excites and interrogates
the sample under study. The spectral response of a sample, which may
arise due to linear or nonlinear absorption may span the entire
OFC spectrum. Such a situation requires a spectrometer to conduct
measurements. Therefore, existing spectrometers have been adapted and 
improved to resolve individual OFC peaks. 

Alternatively, an OFC can be generated using a four-wave mixing 
(FWM) nonlinear-optical process \cite{Boyd}, where, for example,
a laser light at three frequencies $f_1$, $f_2$ and $f_3$ interacts
in a nonlinear-optical medium resulting in a new optical signal
at a fourth frequency $f_4=f_1+f_2-f_3$. If the three original
optical frequencies are a part of a perfectly spaced OFC spectrum,
then the signal at the fourth frequency extends the already
existing OFC spectrum. It is also possible to generate an OFC
using laser light of two equally spaced frequencies, where FWM
can generate light at different equally spaced frequencies via
a cascaded nonlinear process. For example, such an interaction
can produce light at a frequency $2f_1-f_2$ that, in turn, can
subsequently participate in the nonlinear generation of additional
new frequencies in the same OFC spectrum and so forth. This kind
of cascaded OFC generation has been demonstrated in nonlinear
optical fibres \cite{Sef98} and in some nanophotonic devices
\cite{Mak13}.   

Other nonlinear optical processes such as second harmonic generation,
where high-intensity pump light enters a nonlinear optical material
and a weak optical signal is generated at a frequency twice that of
the original pump light, or third harmonic generation of sum and
difference frequency components, or intensity-dependent index of
refraction (Kerr effect) \cite{Boyd} can be employed to create OFC
\cite{Mak19}. In particular, the Kerr effect is used in an important
class of Kerr OFCs or micro-combs \cite{Che16, Wu18, Pas18}, where a
single laser is coupled with a photonic microresonator such as a glass
disc that supports optical whispering-gallery modes. Although such
resonant modes are not exactly equally spaced due to optical
dispersion processes, they can be stabilised, for example, using the
aforementioned FWM effect. Yet, it is noteworthy that, in the time
domain, while mode-locked laser OFCs are virtually always associated
with a series of short pulses, Kerr OFC exhibits complex phase
relations between their individual modes that may not correspond to
well-defined single pulses \cite{Fai16}. However, the modes of Kerr
OFC remain highly coherent thus enabling their application as pure
OFCs.

For some practical applications, OFCs can be generated using an
electro-optical modulation of a continuous wave laser light. Here,
an OFC spectrum is obtained by modulating the amplitude or phase of
a continuous wave laser source using an external modulator operating
at a high radio or microwave frequency \cite{Mur00}. Given this, the
spectrum of a so-generated OFC can be conveniently centred around an
optical frequency of interest. Furthermore, this approach enables
generating OFCs with higher repetition rates of more than 10\,GHz that 
is challenging to achieve using a mode-locked laser \cite{Tor13}. However, 
the number of peaks in the spectrum of electro-optical OFC is lower
than in the spectrum of a mode-locked laser OFC.

Finally, we mention low radio-frequency OFCs generated using purely
electronic and technically simpler devices that produce a series of
pulses. Although such OFCs are used mostly in conjunction with some
functions of electronic sampling oscilloscopes, they have been
utilised in some optical domain applications, for example, in
measurements involving laser diodes and in acoustic frequency 
combs proposed in \cite{Wu19} (Sec.~\ref{sec:3}).

\section{Electronically generated acoustic frequency
  combs\label{sec:3}} 
Marine science has always been of technical, military and commercial
importance. Since about 70\% of the Earth's surface is covered
by water and the global mean sea level is rising due to the climate
change, exploration of oceans has become one of the priorities for
both governmental and private sectors. In particular, in this area
there is an urgent need for novel precision measurement techniques
that would enable the exploration of deep sea also facilitating the
communication and data transfer between submarines and other
equipment. However, in general, optical technologies cannot be used
for these purposes because of a strong absorption of light in
water. On the other hand, the attenuation of acoustic waves in water
is much weaker. This property of sound makes it the prime candidate
for the use in underwater navigation and ranging (SONAR) and other
applications. 

In an idealised underwater distance measurement system, piezo-electric
transducers driven by electric signals produce a train of acoustic
signals and radiate them towards a target. If the speed of sound in 
water is known, the distance to a target is determined by measuring
time between emitted and reflected acoustic pulses. However, the speed
of sound in liquids depends on multiple variable environmental factors
such as the ambient temperature and water salinity
\cite{Uri83}. Therefore, its accurate value may not always be
known. The acoustic measurement accuracy of an unlocked acoustic
device relying on an incoherent data processing method is also
intrinsically limited by destructive interference 
processes and a trade-off between spatial resolution and 
uncertainty in speed measurements, which prevents the current
SONARs from resolving sub-centimetre distances. While such an
inaccuracy is tolerated in some cases, in other applications (for
example, in monitoring underwater glaciers) distance measurement
precision of order of several millimetres is required \cite{Mac17}.

\begin{figure}[t]
  \centering
  \includegraphics[width=.75\textwidth]{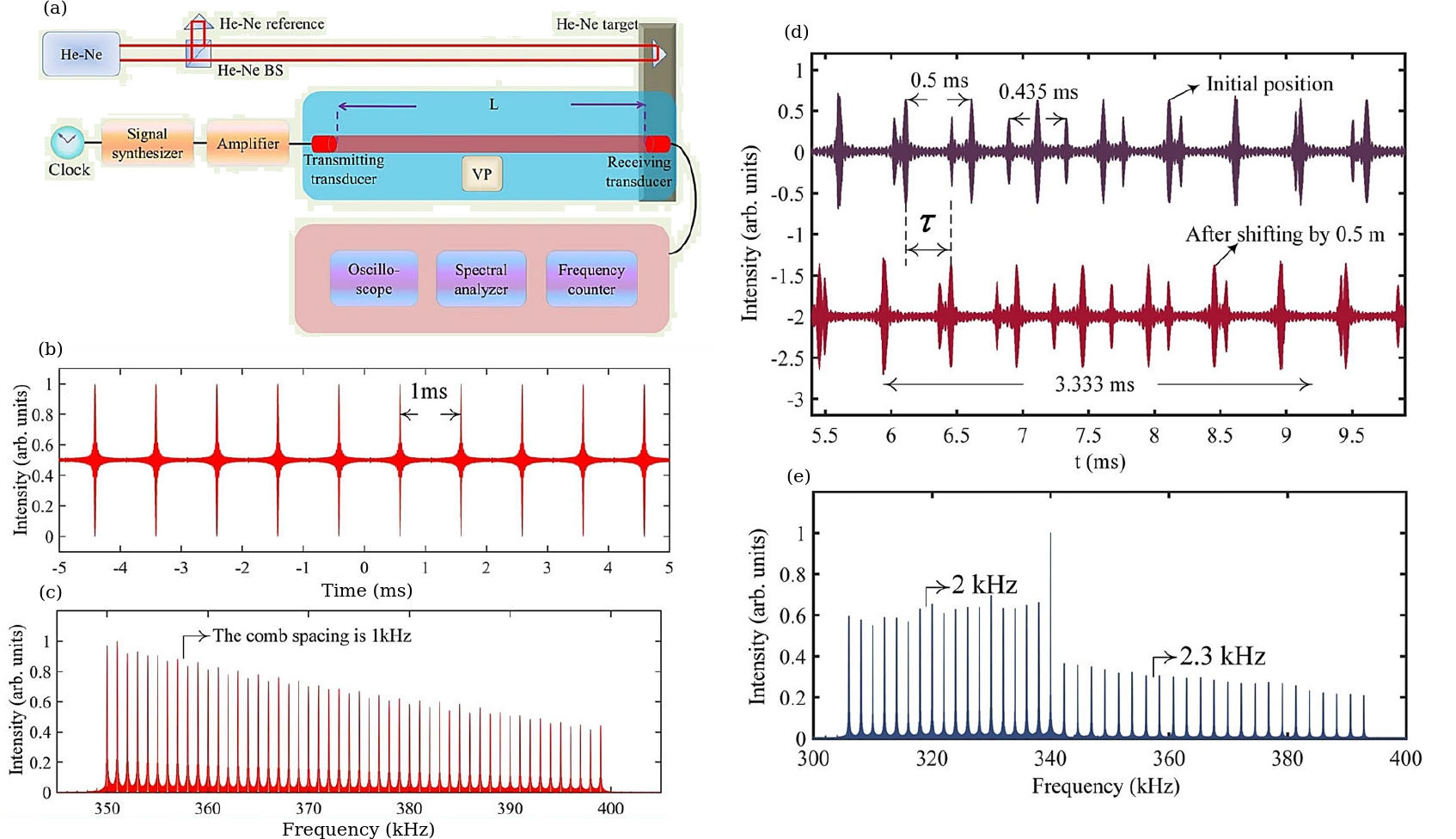}
  \caption{\textbf{(a)}~Schematic of the experimental setup employed
    to generate an AFC using a Rb-clock-stabilised electronic
    FC and a set of piezo-electric transmitting transducers and
    subsequently exploit the resulting AFC for measuring the distance in
    an underwater environment (VP--sound velocity profiler, BS--beam
    splitter).
    \textbf{(b)}~Acoustic pulse train in the time domain in the
    350--399\,kHz frequency range and 
    \textbf{(c)}~the AFC spectrum corresponding to it.
    \textbf{(d)}~Waveforms detected by the receiving transducer in the
    time domain. The upper waveform was detected at the initial
    position and the lower one obtained after moving the receiving
    transducer by 0.5\,m.
    \textbf{(e)}~Spectrum of the waveform in panel~(d) showing two
    different repetition frequencies. Reproduced from \cite{Wu19} with
    permission of John Wiley and Sons.\label{Fig2}}
\end{figure}

Subsequently, a novel approach to underwater distance measurement
was proposed in \cite{Wu19}, where [Fig.~\ref{Fig2}(a)] a signal
produced by a generator referenced to an Rb clock was first amplified
and then used to drive a series of transmitting transducers with
different nominal acoustic frequency bands. A receiving transducers
were fixed at an {\it a priori} known distance, and its output signal
was analysed using an oscilloscope, spectrum analyser and frequency
counter. Since the response of piezoelectrical transducers closely
follows the waveforms of the electric signals used to drive them,
a comb-like spectrum could be generated using a driving signal that
is an electronic FC itself (Sec.~\ref{sec:2}). A He-Ne laser was
used as the reference interferometer in air---the He-Ne laser beam
was aligned with the acoustic beam of the transducers.
Significantly, since the peaks of the electrical FC spectrum are
equally-spaced and fully referenced, the corresponding acoustic 
frequencies emitted by the transducers and the repetition frequency
are as stable as those of a Rb clock [Fig.~\ref{Fig2}(b,~c)]. 

The distance between the transmitting and receiving transducers was
measured using a two-step protocol involving a coarse measurement
followed by a refinement step. During the coarse measurement, the
integer part of the pulse-to-pulse length of the AFC was determined
using its repetition frequency $f_{rep}$. Subsequently, at the refinement
step the fractional part of the pulse-to-pulse length was found using
the slope of the unwrapped phase that is proportional to the time
delay $\tau$ (the phase slope was measured using a Fourier
transform-based approach). Finally, using these data the actual
distance between the transducers was precisely determined.

The maximum unambiguous range of the discussed underwater measurement
system is given by the longest range that a transmitted pulse can travel
forward and back during the time between two consecutive transmitted
pulses. The proposed method of AFC generation can be used to produce
several AFCs at the same time. It has been demonstrated that this can
be employed to expand the unambiguous measurement range. In
particular, in \cite{Wu19} the transmitting transducers were set to
simultaneously emit a pair of AFC signals with two different
repetition frequencies, 2 and 2.3\,kHz. To distinguish between the two
AFCs in they had different amplitudes. Fig.~\ref{Fig2}(d,~e) shows
the detected waveform of the receiving transducer, where the resulting
dual-AFC signals can be seen as the pulse trains with a 0.5\,ms period
(2\,kHz repetition frequency, 0.74\,m pulse-to-pulse length) and
a 0.435\,ms period (2.3\,kHz repetition frequency, 0.640 m pulse-
to-pulse length). A signal with a larger period of 3.33\,ms,
corresponds to the extended unambiguity range of 4.9\,m. Using
this measurement approach, underwater distance measurements up
to 7\,m with stable environmental conditions in an anechoic pool
were conducted with a measurement uncertainty of approximately
50\,$\mu$m compared with the optically-measured reference values.    

\section{Phononic frequency combs\label{sec:4}}
\subsection{Micromechanical resonator-based phononic
  FCs\label{sec:4_1}} 
In this section, we discuss another kind of AFCs, phononic FCs 
\cite{Cao14, Xio16, Cao16, Gan17, Gan19, Kub20, Mer20}, that exploit
high-frequency nonlinear mechanical vibrations \cite{Rud06} to
generate FC-like signals. The cited seminal works have been the first
to demonstrate that the robustness and versatility of the Nobel
Prize-winning OFC technology could be employed in the frequency ranges
that are not accessible using light. Indeed, as shown in \cite{Pic19},
despite a large number of research studies focused on the expansion of
the spectral coverage of the existing OFC generators, there are no
FC-like technologies that would operate in, for example, ultrasound
(MHz) and hypersound (GHz) acoustic frequency ranges
\cite{Fab69}. Although the hypersonic range remains insufficiently
explored compared with audible sound and ultrasound, partly because
GHz acoustic waves undergo a stronger attenuation \cite{Fab69},
hypersound is of a significant technological importance because in
this frequency range one often observes intriguing physical effects
such as Brillouin light scattering (BLS) that originates from a
non-elastic light interaction with acoustic waves
\cite{Gar18}. Similarly to Raman scattering \cite{Tra15}, BLS
underpins an emergent spectroscopy and imaging technique that has
already found important applications in biology, medicine, chemistry,
physics and material science \cite{Bal15, Men16, Pal19, Rem20}. 
We will return to this discussion in Sec.~\ref{sec:5}.

\begin{figure}[t]
  \centering
  \includegraphics[width=12.0 cm]{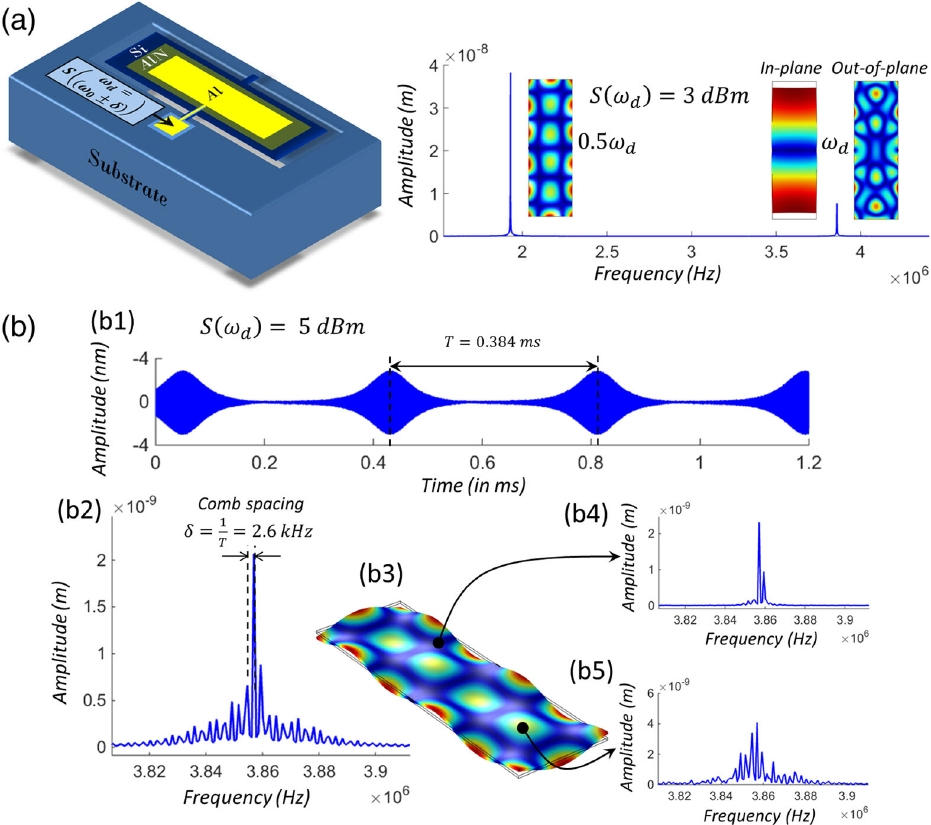}
  \caption{Generation of a phononic FC using nonlinear acoustic  
    resonances in a piezo-electrically driven micromechanical  
    resonator. \textbf{(a)}~Left: Sketch of the micromechanical  
    resonator. Right: Spectrum amplitude plot showing a parametric  
    excitation of an out-of-plane subharmonic mode of the resonator  
    that is tuned on its in-plane extensional mode using a driving  
    signal $S(\omega_d)$=3\,dBm. Panels~(b1) and (b2) show a pulse  
    train corresponding to the FCs generated at $S(\omega_d)$=5\,dBm  
    and its spectrum, respectively. One can see the peaks with an  
    interpeak spacing of 2.6\,kHz. Panel~(b3) shows the displacement  
    profile of the subharmonic mode. Panels~(b4) and (b5) demonstrate  
    that the FC generation is possible only using the displacement at  
    an antinode of the subharmonic mode of the resonator. Reproduced  
    with permission from \cite{Gan17}. Copyright 2017 by the
    American Physical Society. \label{Fig3}}
\end{figure}
Although there are certain physical analogies between phonons and
photons \cite{Mak19}, in general a dispersion relation for phonons is
nonlinear since it is related to acoustic and thermal properties of a
material. As a result, the use of standard methods of OFC generation
in phononic systems is frequently impossible. Therefore, novel
approaches that are independent of the phonon dispersion relation are
required. A solution to this problem was theoretically proposed in
\cite{Cao14} using nonlinear phononic systems such as mechanical
cantilevers \cite{Sad99} and chains of particles linked by springs
that obey Hooke's law but at the same time exhibit a nonlinear
behaviour \cite{Ono15}. In such systems, several phonon modes can be
simultaneously excited by an external driving force producing an
FC-like spectrum with an array of discrete and equidistant spectral
peaks corresponding to the frequencies of nonlinearly excited phonon
modes. FCs generated using nonlinear resonance of different orders
were theoretically investigated in \cite{Cao14} and a possibility of
frequency stabilisation of the higher-order modes was demonstrated.

 Based on the theoretical results presented in \cite{Cao14}, in the work
\cite{Gan17} a phononic FC was experimentally created using a
piezo-electrically driven micromechanical resonator, where an
electromechanical coupling led to signal enhancement and,
consequently, stronger nonlinearities, thereby reproducing the
behaviour of nonlinearly oscillating particle chains considered in
the model proposed in \cite{Cao14}. The micromechanical resonator was
fabricated on a Si chip packaged in a ceramic leadless chip carrier
[Fig.~\ref{Fig3}(a,~left)]. The resonator was driven by electrical
signals produced by a waveform generator and its mechanical response
[Fig.~\ref{Fig3}(b)] was optically recorded by a laser Doppler
vibrometer (LDV) \cite{Mak19}. At a high driving signal amplitude that
exceeded a specific threshold value the data obtained using the LDV
and a spectrum analyser revealed the existence of an
autoparametrically generated subharmonic mode [Fig.~\ref{Fig3}(a,
right)]. The analysis of the displacement profile of the resonator
corresponding to that subharmonic mode [Fig.~\ref{Fig3}(b)]
demonstrated that tuning of the signal registration equipment on an
antinode of the subharmonic mode would be advantageous for the FC
generation. Furthermore, according to the authors of \cite{Gan17}, the
result illustrated in Fig.~\ref{Fig3}(b) speaks in favour of a phase 
coherency of equidistant FC peaks and thus their conceptual analogy with 
the peaks of Kerr OFCs. Since the FC generation was possible only when
the amplitude of the driving signal exceeded a certain threshold
value, further analysis was carried out revealing that when the
driving amplitude was increased, the FC spectrum extended to higher
orders. This important results shows that the spectral bandwidth of
the so-generated FC is directly related to the driving amplitude level.

\begin{figure}[t]
  \centering
  \includegraphics[width=12.0 cm]{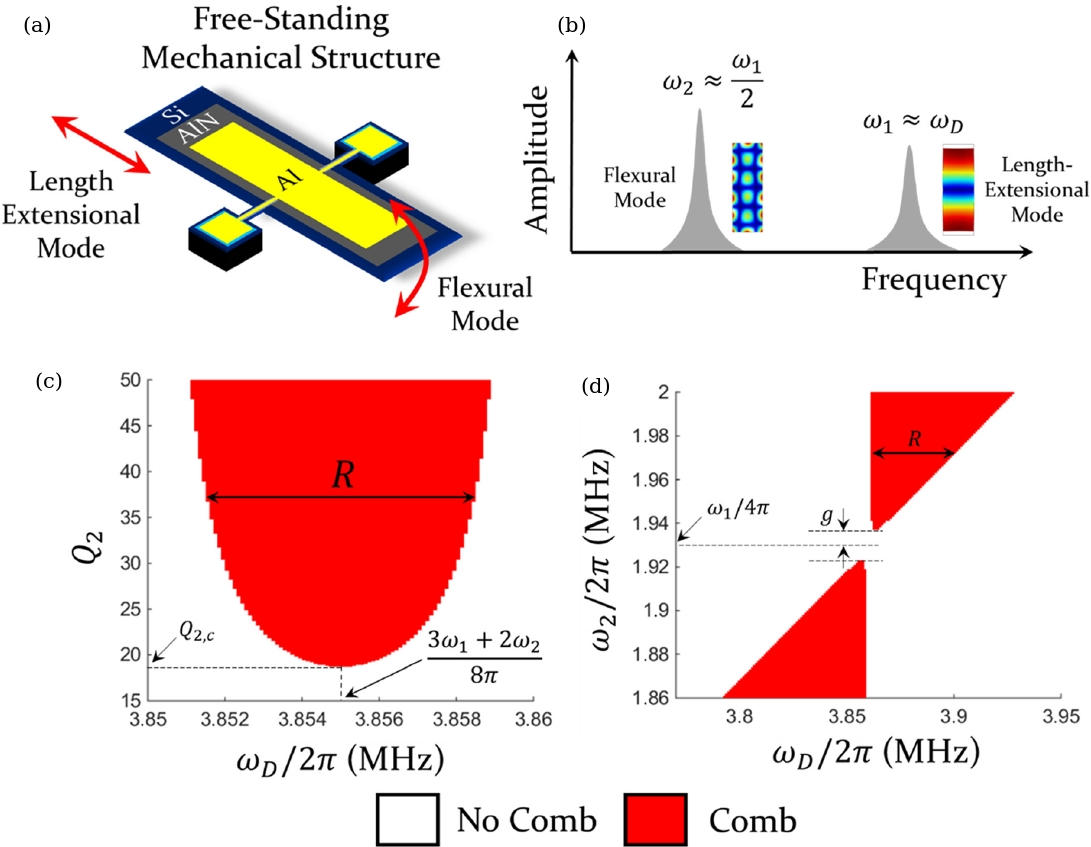}
  \caption{Model of phononic FC generation using nonlinear resonances  
    of a micromechanical resonator. \textbf{(a)}~Sketch of the model  
    of the micromechanical resonator used in the experiment of  
    \cite{Gan17}. 
    \textbf{(b)}~Schematic representation of the experimental spectrum  
    in Fig.~\ref{Fig3}(a, right) showing a fundamental  
    length-extensional mode of the resonator (the resonance frequency  
    $\omega_1$) and its subharmonic flexular mode (the resonance  
    frequency $\omega_2\approx\omega_1/2$.) 
    \textbf{(c, d)}~Graphical representation of the regions, where the  
    generation of FC is possible using the experimental parameters  
    from \cite{Gan17}. Panel~(c) shows a plot of the quality factor of  
    the mode with $\omega_2$ as a function of the driving frequency  
    $\omega_D$. The $\omega_2$-vs-$\omega_D$ plot is presented in  
    Panel~(d). The parameters are $\omega_1/2\pi$=3.86\,MHz and  
    $Q_1$=4000 in (c) and $\omega_1/2\pi$=3.86\,MHz, $Q_1$=4000 and 
    $Q_2$=50 in (d). Reproduced from \cite{Qi20} with the permission
    of AIP Publishing. \label{Fig4}}
\end{figure}

 In the follow-up work \cite{Qi20}, the influence of the phonon mode
structure on the FC generation was investigated using a model of two
nonlinearly coupled phonon modes [Fig.~\ref{Fig4}(a, b)]. The model
predicted the existence of a region within the amplitude-frequency
space where the FC generation is possible [Fig.~\ref{Fig4}(c, d)]. The
frequency range $R$ corresponding to this region is given by the
expression
$R=\left|\omega_2-\dfrac{\omega_1}{2}\right|-\dfrac{\sqrt{2}\omega_1}
{\sqrt{Q_1Q_2}}$, where $\omega_1$ is the resonance frequency of the
fundamental length-extensional mode of the resonator, $\omega_2$ is
the frequency of its subharmonic flexular mode and $Q_1$ and $Q_2$ are
the respective quality factors [Fig.~\ref{Fig4}(a, b)]. One can see
that $R \to\left|\omega_2-\frac{\omega_1}{2}\right|$ for large values of
$Q_2$. There also exists a critical value for
$\left|\omega_2-\dfrac{\omega_1}{2}\right|$ given by parameter
$g=2\omega_1\sqrt{\frac{2}{Q_1 Q_2}}$, which implies that for
$\left|\omega_2-\dfrac{\omega_1}{2}\right|>g$ the frequency range of
the FC existence $R$ scales linearly with
$\left|\omega_2-\dfrac{\omega_1}{2}\right|$ [Fig.~\ref{Fig4}(d)].

It was found that the region of FC existence originates from a subset
of the Arnold tongues \cite{Rab89}, a phenomenon known in the context 
of the interaction between oscillators, where one oscillator drives
another. In particular, Arnold tongues have been observed in a
two-oscillator system, where one oscillator influences the other but
not vice-versa, which is typical of oscillators driven by a periodic
force. Moreover, it was established that the spectral location and
composition of the region of FC existence can be analytically defined
in terms of resonance frequencies, quality factors and mode coupling
strength of the mechanical resonator as well as by a detuning of the
driving frequency from those of the mechanical resonances. 

It is also noteworthy that the FCs discussed in this section have not
been precisely stabilised to a frequency reference. Therefore,
similarly to Kerr OFCs, they cannot be considered as counterparts of
mode-locked laser OFC (Sec.~\ref{sec:2}). This fact was perceived as a
fundamental limitation for applications such as underwater distance
measurements \cite{Wu19} because the modes of mechanically-generated
FCs could be incoherent. Nevertheless, the results discussed in this
section have opened novel opportunities in the fields of
ultrasensitive vibration detectors \cite{Kum14}, phonon lasers
\cite{Gru10, Bea10}, quantum computers \cite{Sta12} and imaging
\cite{Cao14, Gan17, Mak19}, where some incoherence of FCs can be
inconsequential. 
 
\subsection{Phononic frequency combs in bulk acoustic wave systems} 
The discussion of mechanically-generated FCs in Sec.~\ref{sec:4_1}
demonstrates that to achieve an FC-like signal in such systems one
needs to design a certain mode structure and to apply a high threshold
driving force. While these requirements can be fulfilled in many
practical situations, their realisation may be impossible in some
applications. For example, this is the case for a large group of 
technologies that include stabilised low-noise classical and atomic
oscillators and measurement systems, high-sensitivity displacement
sensors, high-precision electron spin and ferromagnetic resonance
spectroscopy, high-precision measurement of material properties and
high-quality-factor hybrid quantum systems \cite{Mak18, Gor20, Bai21}.
The aforementioned technologies enable the realisation of precision
measurement tools and techniques to test some of the core concepts of
fundamental physics, such as modern searches for Lorentz invariance
violations in the photonic \cite{Nag15}, phononic \cite{Lo16, Gor18_1}
and gravity domains \cite{Sha19}, variations in fundamental constants
\cite{Gue12} and research on dark matter \cite{Gor19}. In these
applications, bulk acoustic wave (BAW) devices have found
very extensive applications. Moreover, both bulk and surface
acoustic wave (SAW) devices have been used for spectroscopy,
detection and sensing \cite{Joh08}. Nonlinear dynamics of
BAW and SAW mechanical systems, including FC generation, has also 
become a subject of theoretical research \cite{Has13, Roq20}. 

\begin{figure}[t]
  \centering
  \includegraphics[width=12.0 cm]{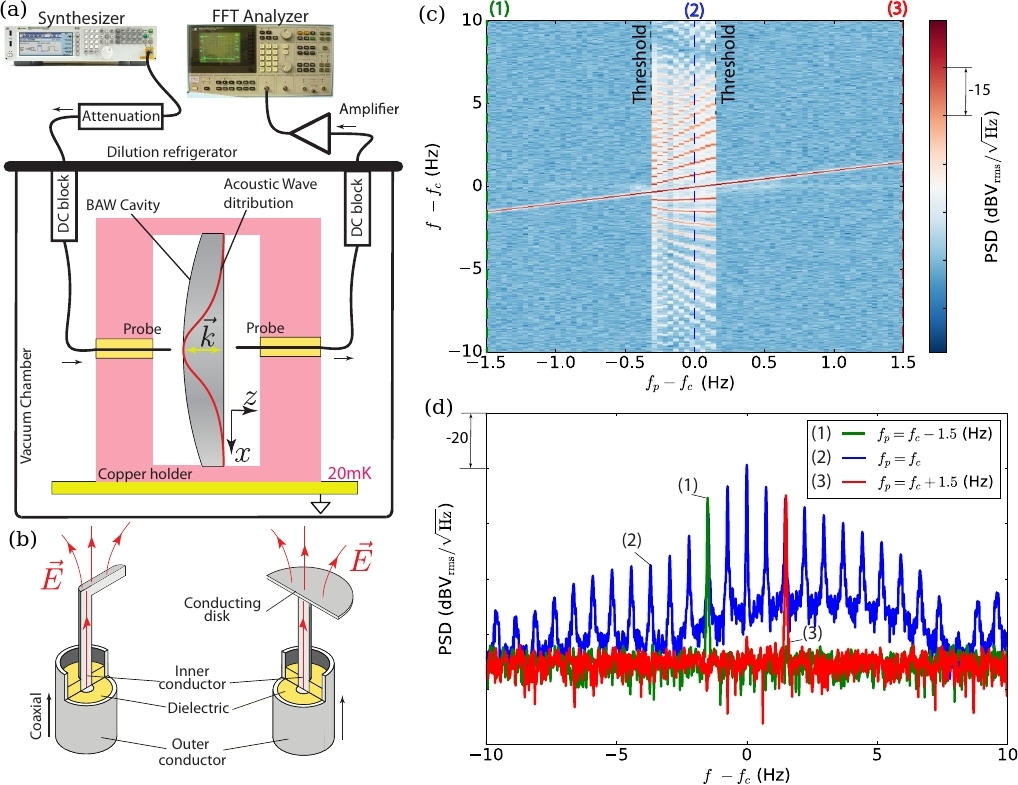}
  \caption{Generation of a BAW-based phononic FC. \textbf{(a)}~Sketch  
    of the experimental setup showing the BAW resonator, copper holder  
    and excitation probes placed in a vacuum chamber.  
    \textbf{(b)}~Illustration of two types of excitation probes used  
    in the experiment:  L- (left) and disc-shaped (right)  
    antennae.  
    \textbf{(c)}~A false-colour map composed of the individual output  
    signal PSDs $S(f-f_c)$ plotted as a function of the pump frequency  
    detuning $f_p-f_c$ at a constant incident power $P=-66$\,dB\,m.  
    \textbf{(d)}~Three PSD curves for different constant incident  
    frequencies $f_p$ and the same power $P$ corresponding to slices 
    (1), (2) and (3) along the vertical axis of the plot in  
    Panel~(d). Reproduced from \cite{Gor20} published by the American
    Physical Society under the terms of the Creative Commons Attribution
    4.0 International license. \label{Fig5}}
\end{figure}

Recently, the generation of phononic FCs in a BAW system at a
temperature of 20\,mK using a single-frequency low-power signal source 
was demonstrated \cite{Gor20}. To enable such a generation, in general
one needs a system with low losses and strong nonlinear effects. To
achieve this, a phonon-trapping stress compensated quartz BAW cavity
operating at 20\,mK was employed [Fig.~\ref{Fig5}(a)]. Quartz BAW
cavities are known to have extremely high values of quality factors at
cryogenic temperatures reaching $Q=8\times10^9$ \cite{Gal13}. At the
same time, quartz BAW cavities possess significant 
acousto-mechanical nonlinearities that originate from the lattice
non-harmonicity \cite{Tie76, Nos99} that can be described using
Duffing oscillator---a nonlinear second-order differential equation
used to model certain damped and driven oscillations \cite{Rab89}.
There are also other sources of nonlinearity originating from
thermoelectroelastic effects and physical processes, which
are observed mostly at milli-Kelvin temperatures, such as
coupling with ensembles of two level systems (TLSs) \cite{Lis15}.
In particular, the presence of TLSs has been demonstrated in BAW
cavities through a number of effects including nonlinear losses and
magnetic hysteresis. However, these processes hinder the generation of
FCs and should be avoided.

This approach was adopted in the experiment reported in \cite{Gor20},
where the BAW cavity was placed inside a copper holder attached
to a base plate of a refrigerator cooled to 20\,mK. To ensure
that no coupling exists between the holder and BAW the fundamental
microwave resonance frequency of the former was chosen to be much
higher than any resonance frequency of BAW. The acoustic modes
were excited and detected piezoelectrically using two specially
designed coaxial probes coupled to the electric field of the
structure [Fig.~\ref{Fig5}(b)].

In one of the experiments reported in \cite{Gor20}, the system
response was analysed in terms of the signal power spectral density
(PSD) $S(f)$ measured as a function of the pump signal frequency $f_p$
in the vicinity of the acoustic resonance $f_s$ at a constant
signal power $P$, see Fig.~\ref{Fig5}(c), where a false-colour
map is composed of individual PSD curves of the output signal for
each incident signal frequency $f_p$ used in the experiment. 
Note that each PSD curve was obtained independently
with a time delay between two consecutive measurements sufficient
to suppress any residual signals coming from a preceding measurement.
It is also noteworthy that the experimental setup was stable on
the time scale of the measurements since it was locked to an
atomic frequency standard, and that the characteristics of the
BAW resonator were found to be insensitive to possible temperature
fluctuations during the measurement.

Figure~\ref{Fig5}(d) shows three PSD curves obtained for different
values of the incident signal frequency $f_p$ at a constant power $P$
corresponding to slices (1), (2) and (3) along the vertical axis
in Fig.~\ref{Fig5}(c). A FC is generated when the pump signal
frequency approaches the resonance frequency. Furthermore, the FC
exhibits two thresholds on each side of the resonance and the FC
repetition rate is about 0.8\,Hz when $f_p = f_c$. In the subsequent
experiments reported in \cite{Gor20} the same threshold of the FC
generation was observed when the incident power was varied but the
excitation frequency was $f_p\approx f_c$. The analysis of the
experimental results also revealed that the FC spectrum significantly
depends on geometry of excitation and detection electrodes
Fig.~\ref{Fig5}(b). Yet, the fact that a strong Duffing nonlinearity
was observed below the generation threshold indicates that the 
system is a phononic analogue to Kerr OFCs excited in monolithic
optical microresonators (Sec.~\ref{sec:2}). Thus, it was concluded 
that the ultralow power regime explored in \cite{Gor20} opens a way
for integrating BAW-based phononic system with a quantum hybrid
counterpart such as superconducting qubits.
 
\section{Brillouin light scattering-based frequency
  combs\label{sec:5}}
In this section, we discuss the recent achievements in the developing
field of FCs generation using Brillouin light scattering (BLS), a
physical effect named after L{\'e}on Brillouin, where light interacts
with material waves in a medium \cite{Fab69}. Such an interaction
is enabled by a dependence of the optical refractive index on the material
properties of the medium. For example, it is well-established that
the refraction index of a transparent material changes when it is
mechanically deformed. As a result of a deformation, a small fraction
of light that is transmitted through the material or reflected from it
changes its momentum (i.e.~its frequency and energy are changed). This
process is similar to an effect, where diffraction of light is caused
by diffraction grating the components of which vibrate with a
frequency that is much smaller than the frequency of the light
wave. In solid media, macromolecular aggregates, biological 
media and liquids and gases, BLS can be observed as a result of light 
interaction with acoustic (phononic) modes \cite{Fab69, Bal03, Pal19,
  Men16, Tra15}, exciton-polariton (a hybrid light and matter
quasiparticle arising from a strong coupling of the electromagnetic
dipolar oscillations and photon) \cite{Pod14} and spin waves and their
quanta---magnons \cite{Mak15}---existing in magnetic materials
\cite{Dem01, Sta07, Gub10, Ser12, Seb15, Mak15_review}.

Although Rayleigh scattering can also be considered to be due to
fluctuations in the density of an optical medium, thus leading
to variations in its refraction index, such fluctuations are of
random and incoherent nature. In contrast, BLS is caused by correlated
periodic fluctuations such as phonons and magnons. Therefore, Rayleigh
scattering involves no energy loss or gain. On the other hand, although
Raman scattering also involves inelastic interaction processes caused by
vibrational properties of matter, the range of frequency shifts associated
with this effect are very different compared with those in BLS. Thus,
BLS and Raman scattering provide very different information about
the sample under study: Raman spectroscopy enables one to determine
the chemical composition and molecular structure of the medium while
BLS senses the elastic properties of the material \cite{Men16, Aki18}.

\begin{figure}
  \centering
  \includegraphics[width=12.0 cm]{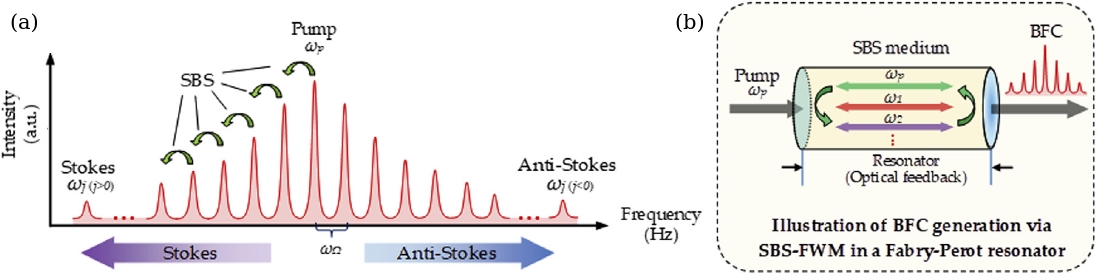}
  \caption{Illustration of the generation of an SBS-based FC.  
    \textbf{(a)}~The spectrum of an SBS FC showing the Stokes 
    and anti-Stokes frequency peaks generated due to SBS effect.  
    \textbf{(b)}~Illustration of a cascaded SBS effect combined with  
    FWM arising from the optical Kerr-nonlinearity. Reproduced from
    \cite{Bai18} with permission of Elsevier. \label{Fig6}}
\end{figure}
 Before we discuss FCs generated using BLS, we also note a conceptual
difference between this effect and stimulated Brilouin scattering
(SBS) \cite{Gar64, Bai18, Gar18}. SBS arises when an intense beam of
laser light propagates through an optical medium, such as an optical
fibre, and when variations in the optical electric field of the beam
itself induce acoustic vibrations in the medium via electrostriction
and radiation pressure effects. Under these conditions, the beam may
display BLS as a result of the interaction with the vibrations
leading to the generation of optical signals with a spectrum
consisting of a large number of equally spaced peaks that are
coherently phased---a Brillouin OFC (BFC) [Fig.~\ref{Fig6}]
\cite{Bra09, Lin14, Lu16, Don16, Bai18, Egg19}. However, the 
so-generated FCs belong to the group of OFCs since SBS is mostly
a nonlinear-optical effect that was discovered only after the invention
of a laser \cite{Gar18, Bai18, Boyd}. Therefore, we refer an interested
reader to the cited studies and references therein while in the
following we focus on FCs based on the original BLS effect.

\subsection{Magnonic BLS-based frequency combs}  
While BLS offers significant advantages in several research fields, in
general this technique requires sophisticated experimental
instrumentation to reliably detect light scattering frequency
shifts. A typical frequency shift observed in BLS measurements ranges
from several MHz to several GHz, which is a very small compared to the
frequency of the incident light (several hundreds of THz). As a
result, in an optical spectrum the BLS peaks are located on the
shoulders of the central Rayleigh scattering peak and their amplitudes
are so small that to resolve them a Sandercock multi-pass
Fabry-P\'erot interferometer \cite{Moc87} or a virtual-image phase
array (VIPA) spectrometer has to be used \cite{Sca08}. However,
despite these technical challenges, BLS spectroscopy has been an
essential tool for research on spin wave excitation in ferromagnetic
micro- and nano-structures \cite{Dem01, Sta07, Gub10, Ser12, Seb15,
  Mak15_review} and on phonon excitations in solid and biological
media \cite{Fab69, Bal03, Pal19, Men16, Tra15, Aki18}.

\begin{figure}
  \centering
  \includegraphics[width=12.0 cm]{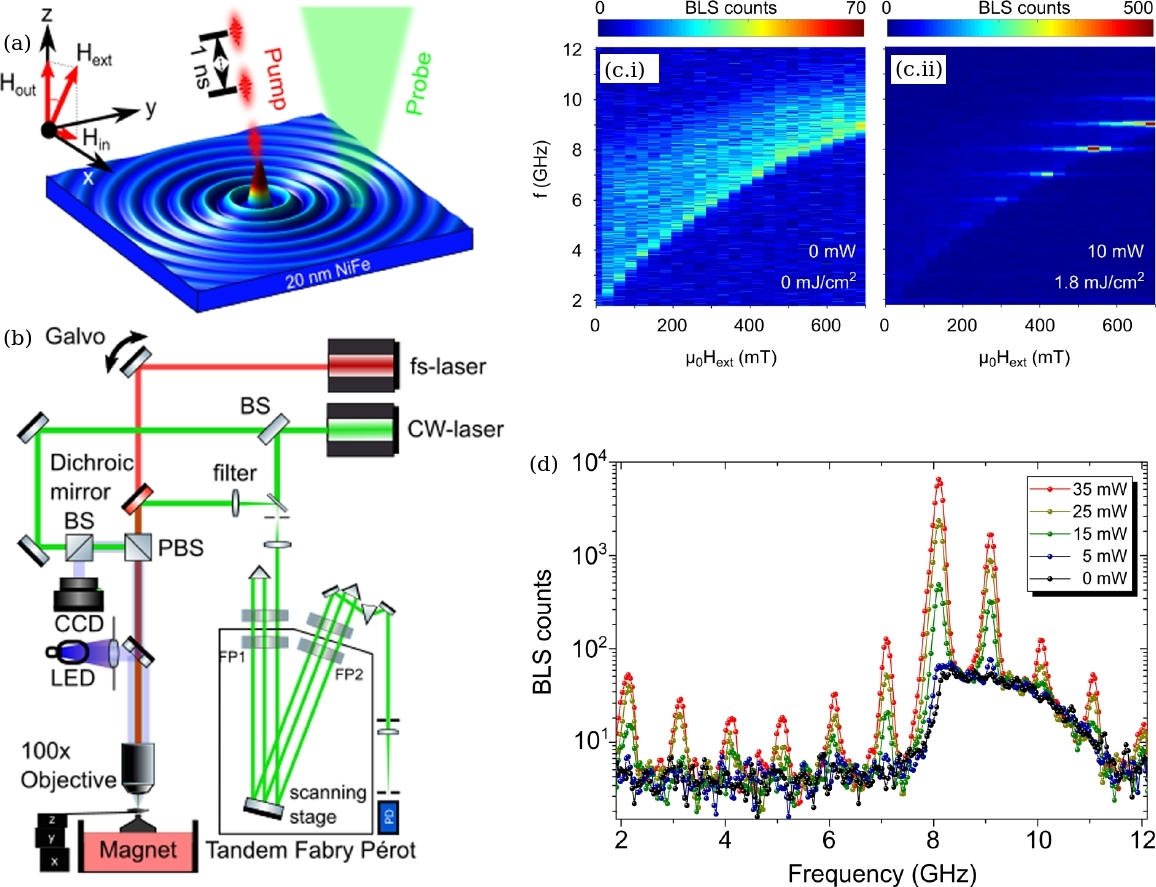}
  \caption{Illustration of the generation of a magnonic BLS-based FC.  
    \textbf{(a)}~A ferromagnetic thin film sample is pumped with a red  
    (816\,nm wavelength) 120\,fs-long pulsed laser light and probed  
    using a continuous green laser (532\,nm) light. The direction of  
    the external magnetic field $H_{ext}$ applied to the sample is  
    indicated. 
    \textbf{(b)}~Schematic of the optical setup of the pump-probe  
    experiment, where BS denotes a 50/50 beam splitter and PBS is a  
    polarising beam splitter. The sample is placed right below a  
    microscope objective to achieve diffraction-limited focusing. The  
    backscattered light is analysed using a six-pass tandem  
    Fabry-P\'erot interferometer (TFPI) and detected using a  
    single-channel avalanche photodiode (APD).  
    \textbf{(c)}~False-colour maps of thermally excited spin wave  
    spectra as a function of the applied magnetic magnetic field  
    without (c.i) and with (c.ii) irradiation with a fs-laser pulse  
    train at a 1.8\,mJ/cm$^2$ fluence.  
    \textbf{(d)}~BLS counts as a function of frequency at an applied  
    magnetic field magnitude of 600\,mT for four different laser  
    powers. The thermal spin wave background is also shown. Reproduced  
    from \cite{Mur20} published by the American Physical Society under
    the terms of the Creative Commons Attribution~4.0 International
    license.\label{Fig7}}
\end{figure}

Recently, the BLS spectroscopy has been employed to produce FC-like
signals originating from spin wave modes excited in a ferromagnetic thin
film structure \cite{Ale20, Mur20}. In those works, a thin Permalloy
(Ni$_{80}$Fe$_{20}$ alloy) film---a standard building block of many
magnonic and spintronic devices \cite{Mak15}---was deposited onto a
sapphire substrate using a dc magnetron sputtering technique. A
sapphire substrate was chosen due to its negligibly small optical
absorption at the frequency of the laser light used in the BLS setup
and also due to its high thermal conductivity. The fabricated films
were characterised using a pump-probe experimental setup schematically
shown in Fig.~\ref{Fig7}(a, b), where the pump beam was emitted by a
mode-locked laser with a 1\,GHz repetition rate at the wavelength of
816\,nm with a 30\,fs pulse duration and pulse energies of up to
1\,nJ. Since the emitted laser pulse stretches during the propagation
in the optical system, its actual duration at the moment of time when
it reaches the sample was 120\,fs. A microscope objective was used to
focus the laser beam into a spot with the size approaching the optical
diffraction limit (approximately 400\,nm). The magnetisation 
dynamics in a Permalloy film was probed using a single-frequency
532\,nm laser light that was also focused to the diffraction limit
using the same microscope objective. The probe light scattered backwards
from the sample was first collected and filtered using a polariser and
then analysed using a six-pass tandem Sandercock Fabry-P\'erot
interferometer and then detected using a single channel avalanche
photodiode. The pump laser beam was scanned over the sample using a
pair of galvanometer mirrors and lenses that enabled changing the
lateral distance between the pump and the probe beams.

 Figure\,\ref{Fig7}(c.i) shows the typical field dependence of the
thermal spin wave spectrum obtained for a thin Permalloy film
sample when the pump laser was turned off. In this figure a sharp
cutoff of the spin wave band is seen that is in good agreement 
with the frequency-vs-applied magnetic field dependence theoretically
predicted by Kittel equation \cite{Mak15_review}. The corresponding
spin wave spectrum at the same location of the Permalloy film but
with the pump laser turned on is shown in Fig.~\ref{Fig7}(c.ii).
In this case, the spectrum dramatically changes its character by
gaining a number of distinct equally spaced peaks that appear with the
1\,GHz repetition rate. A more detailed information about the
generated FC-line spectrum is given in Fig.~\ref{Fig7}(d) that shows
the BLS counts as a function of the frequency at an applied magnetic
field of 600\,mT for four different laser powers, also showing the
thermal spin wave background. From this result, one can deduce that
5\,mW is the threshold power level for the FC generation. 

Note that studies reported in \cite{Ale20, Mur20} did not aim
to generate FC signals for the use in applications, where other kinds 
FCs have been typically used. Instead, a method of FC-enhanced BLS
microscopy was introduced to coherently excite vibrational and spin
wave modes in the sample. This new approach is more advanced than a
conventional impulse-driven stimulated BLS, where the spatial resolution
is limited by the size of a virtual grating induced in the medium by
the pump laser beam. Nevertheless, those results are of immediate
relevance to the mainstream discussion in the current review article
since they facilitate the development of BLS-based FC techniques
and promote a deeper understanding of fundamental physical processes
that underpin their operation. These include the enhancement of weak
BLS signals using surface plasmon resonances supported by metal thin
films, gratings and nanostructures \cite{Mak15_review, Mak16, Mak18}
that we discuss in the subsequent sections.

\subsection{Plasmon-enhanced Brillouin light scattering effect}
Surface plasmons are optical waves that propagate along a
metal-dielectric interface \cite{Raether, Eno12}. A localised surface
plasmon mode is a special case of a surface plasmon wave that is
confined to a metal grating or a nanoparticle. Typically, to create
conditions favourable for the excitation of localised plasmons at
least one dimension of the structure supporting them must be
comparable with or smaller than the wavelength of the incident
light. For example, localised plasmons have been observed in spherical
nanoparticles with the diameter of 10--50\,nm and in nanorods that are
50--200\,nm long and typically have a diameter of about 20\,nm. The
optical properties of localised plasmons have been used to enable many
essential operations of light control and manipulations at the
nanoscale. For example, localised plasmons have been exploited to 
dramatically enhance the local optical electric field and then use it
to enhance the light-matter interaction processes that are essential
for achieving strong nonlinear effects \cite{Kau12, Pan18} and
high sensitivity \cite{May11} in many practical situations.

Usually, nanoparticles and nanostructures supporting plasmon modes are
made of gold or silver because these two metals exhibit relatively low
absorption losses at the optical frequencies. Nevertheless, in many
practical situations gold and silver are combined with or substituted
by ferromagnetic metals such as nickel, cobalt, iron and their alloys
(e.g.~Permaloy \cite{Kos13}). Although the absorption losses in plasmon
structures made of these materials can be even higher than in devices
made of pure gold or silver, the ferromagnetic metals and their alloys
exhibits a significant magneto-optical activity \cite{Zvezdin}, which
opens up avenues for ultra-fast control of light and high-sensitivity
biosensing thus laying a foundation of the field of magneto-plasmonics 
\cite{Bon11, Che11, Che12, Tem12, Arm13, Chi13, Mak15_review, Mak16}.

Advances in magneto-plasmonics are also relevant to the current
discussion of BLS-generated FCs. Indeed, in a typical magnonic BLS
experiment, the dispersion relationship of spin waves is determined by
using $p$-polarised incident monochromatic light illuminating the
sample at an angle $\theta$ that is linked to the wave vector $k_{SW}$
of the probed spin wave via the relationship
$k_{SW}=(4\pi/\lambda)\sin{\theta}$, where $\lambda$ is the wavelength
of the incident light. Note that the same polarisation of the incident
light is required for the excitation of surface plasmons in thin films
and grating \cite{Raether} including those made of Permalloy
\cite{Kos13, Mak15_review} (such grating is called magnonic crystals
in magnonics). Subsequently, by analogy with the plasmon-enhanced
magneto-optical response, the conditions for a resonant enhancement of
the BLS signal due to surface plasmons can be satisfied in BLS
measurements of ferromagnetic metal structures.

While this approach has not yet been validated experimentally,
experimental evidence speaks for its plausibility. In fact, in
magnonic BLS experiments the interaction of light with magnetic modes
in the sample is mediated by the magneto-optical effects such as the
magneto-optical Kerr effect (MOKE) and Faraday effect \cite{Zvezdin,
  Dem01, Che12, Chi13, Arm13}. Here, MOKE represents a change 
in the polarisation and intensity of light that is reflected from the
surface of a magnetised material. Similar to Faraday effect, MOKE
originates from the off-diagonal dielectric permittivity tensor
components of the investigated magnetised material
\cite{Zvezdin}. However, while Faraday effect is observed in
transmission and, consequently, occurs only in optically transparent
materials, the observation of MOKE is possible mostly in highly
optically-reflecting samples. Given this, MOKE has been found to be
especially suitable for studying magnetism of highly reflective
metals. Since plasmon modes can be enhanced in the same metal
structures, where MOKE is observed, it has been shown that the
amplitude of the MOKE signal can be increased using both surface and
localised plasmon waves \cite{Bon11, Tem12, Arm13, Kos13, Che12,
  Chi13, Mak15_review}. Thus, since MOKE is also the main physical
mechanism contributing to BLS and since its strength can be
increased using a magneto-plasmonic technique, it is plausible that the
BLS signal would also be amplified by a plasmonic interaction.  

Another significant argument in favour of this assumption is an experimental
demonstration of a plasmon-enhanced phononic BLS \cite{Men15}. While
phononic BLS is conceptually similar to the magnonic one, physical
processes that underpin it are simpler compared with the complex
magneto-optical effects, which facilitates magnonic BLS. From the
experimental point of view, realisation of a BLS experiment involving
a measurement of plasmonic enhancement is technically simpler in
phononic BLS than in magnonic one since a magnetic field needed to
magnetise a sample in the latter case is not required (such a field is
often created by bulky and expensive electromagnets that can also
obstruct the sample from the source of light and photodetectors
receiving it \cite{Kos13}). 

\begin{figure}
  \centering
  \includegraphics[width=12.0 cm]{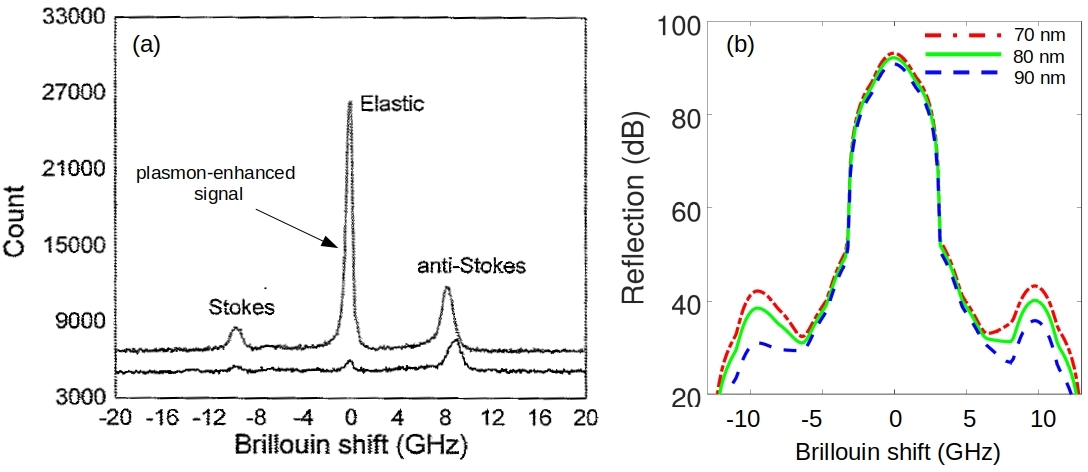}
  \caption{\textbf{(a)}~Typical experimental BLS spectra with and  
    without a gold disc nanostructure obtained after 5\,s   
      integration time using the excitation laser power of  
    40\,mW. Note a significant signal enhancement due to the plasmon  
    modes supported by the disc nanostructure. Reproduced from \cite{Men15}
    with permission of SPIE and the corresponding author of this publication.  
    \textbf{(b)}~BLS spectra calculated using a 3D finite-difference  
    time-domain model of the plasmon BLS effect for different disc  
    diameters. Due to high computational demands, the integration time  
    in the model was limited to 0.5\,ms and window filtering was used,  
    which increased the linewidth of the peaks. Note that the model  
    reproduces the fact that in the experiment the amplitude of the  
    anti-Stokes peak is higher than that of the Stokes peak. The BLS  
    signal without the nanodiscs was not calculated. \label{Fig8}}
\end{figure}   
 A novel approach resolving the issue of a low strength of the BLS
signal and enabling one to overcome such drawbacks of BLS
spectroscopy as a long acquisition time and poor spectral resolution
was proposed in \cite{Men15}. There the enhancement of BLS at the
light wavelength of 532\,nm was investigated using various acoustic
modes of an alkaline-earth boroaluminosilicate glass plate coated with
periodic arrays of gold nanodiscs that support localised surface
plasmon modes. A similar enhancement was also observed from the bulk 
phonons, when the gold nanodiscs were covered by liquids such as
methanol and water. The observed enhancement [Fig.~\ref{Fig8}(a)] was
attributed to the excitation of a fundamental plasmon mode of the
array of nanodisc, which was confirmed by the fact that no enhancement
was observed without the nanostructure and that the enhancement of BLS
was a function of the nanodisc aspect ratio and diameter. It was
suggested that the demonstrated plasmonic enhancement could be
combined with the virtually imaged phased array (VIPA) based
background-free BLS spectroscopy to optimise the acquisition time, and
that an array of nanodiscs could serve as a platform for a practical
implementation of surface-enhanced BLS technique analogous to the 
well-established surface-enhanced Raman spectroscopy (SERS)
\cite{Tra15, Li18, Li19}.
 
To further investigate the origin of the plasmon-enhanced BLS process, we
numerically modelled the BLS interaction in nanodiscs covered by water
using a finite-difference time-domain method \cite{Mak16_hydro}.
Figure\,\ref{Fig8}(b) shows the calculated BLS spectra for three different
disc diameters that were used in experiments reported in \cite{Men15}.
The model reproduces a plasmonic enhancement of the BLS signal and
demonstrates that the amplitude of Stokes and anti-Stokes peaks
depends on the nanodisc geometry. Note that the BLS signal was not 
modelled without a nanodisc structure and that in the model the integration
time (0.5\,ms) was much smaller than in experiment (5\,s) due to
computational constraints. Nevertheless, the model was able to
reproduce the experimental fact that the amplitude of the anti-Stokes
peak is higher than that of the Stokes peak. This speaks in favour of its 
physical veracity. Note also that the linewidth of the peaks in the
spectra in Fig.~\ref{Fig8}(b) is artificially broadened due to the use
of a window filter while postprocessing the simulated BLS
signals. However, even though this artefact has complicated the
analysis of the impact of the plasmon enhancement on the peak
linewidth, a close inspection of the peaks reveals the signs 
of a dependence of both linewidth and Brillouin shift on the nanodisc
geometry, which was also observed experimentally \cite{Men15}.

\begin{figure}
  \centering
  \includegraphics[width=12.0 cm]{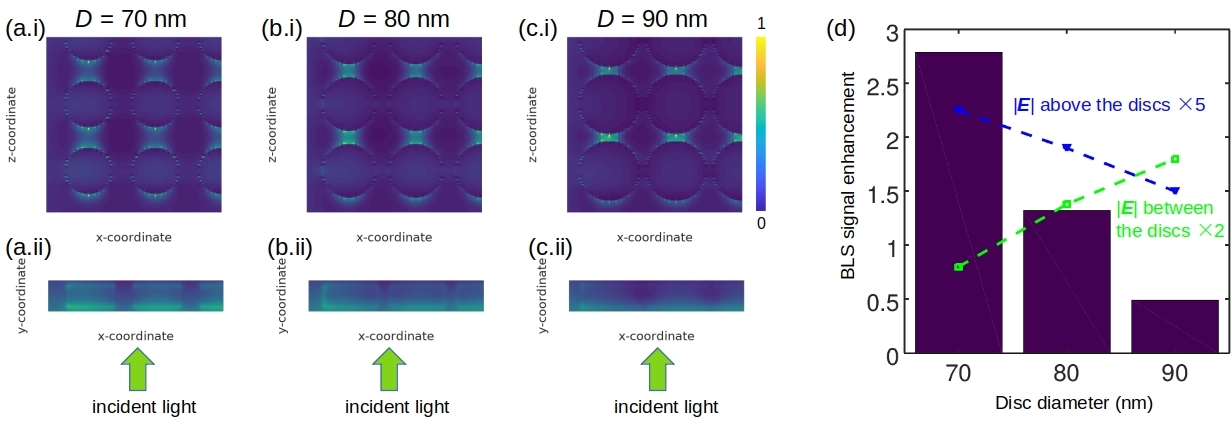}
  \caption{\textbf{(a)}~Top and side views of the simulated optical  
    electric field amplitude |{\bf E}| of the plasmonic modes in the  
    nanodisc array with different disc radii $D$. The direction of 
    propagation of the 532\,nm-wavelength incident light is indicated.  
    \textbf{(b)}~Simulated BLS enhancement as a function of the disc  
    diameter $D$. The triangles and squares and the straight lines  
    (guides to the eye) corresponding to them show the dependence of  
    the field enhancement above and between the discs, respectively. 
    \label{Fig9}}
\end{figure} 
To gain a further insight into the origin of the plasmon-enhanced 
BLS, we simulated the scenario of a 532\,nm wavelength light that is
normally incident on a 5$\times$5 nanodisc array covered by water and
that is polarised along the $z$-axis of the coordinate system adopted
in Fig.~\ref{Fig9}(a--c). In this model, the symmetry of the disc
array was used to reduce computational effort. The spacing between the disc
centres was fixed at 100\,nm while their diameter $D$ was varied.
Figure~\ref{Fig9}(a--c) shows that the enlargement of the disk diameters
slightly enhances the amplitude of the local optical electric field
|{\bf E}| in the gaps between them. However, this effect plays an adverse
role in achieving a stronger BLS response since the light localised 
in the gaps also penetrates the metal surface of the discs, which
results in higher absorption losses. To demonstrate this, in
Fig.~\ref{Fig9}(d) we use bars to represent the BLS enhancement
obtained from the anti-Stokes peak of the simulated spectra. The blue
line connecting triangles shows the dependence of the field
enhancement on the disc diameter calculated immediately above the
discs, i.e.~in the areas, where light senses the refractive index
modulation induced by phonons in the water layer. The green line
connecting squares depicts the field enhancement in the gaps between
the discs. We conclude that plasmonic enhancement of the BLS signal is
possible when light is localised above the discs and not in the gaps
between them. Conversely, the BLS signal is reduced when light is
localised between the discs. This effect is especially pronounced in
the case of large disc diameters $D=90$\,nm and very small 10-nm-wide
gaps between the discs, where our simulations predict a sharp decrease
in the amplitude of the BLS signal compared with the predictions made
for the arrays of nanodiscs with smaller diameters.

\subsection{Application of plasmon-enhanced BLS in frequency comb
  generation} 
As follows from the discussion in the previous section the nanodisc
geometry is not optimal for enhancing the BLS signal. Therefore,
further analysis was performed in \cite{Mak16_hydro}, where it was
shown that a stronger plasmonic enhancement of the BLS effect could be
achieved using elongated metal nanostructures such as plasmonic 
nanorods made of gold or silver. Similar to the well-known Fabry-P\'erot 
resonators, long nanorods can support higher-order plasmonic modes
(those modes do not exist in short nanorods that support only a
dipole-like fundamental mode similar to the fundamental mode of
nanodiscs). The operation based on a higher-order mode is expected to 
be advantageous for enhancing BLS signals since such modes reflect
from a nanorod ends multiple times. This effectively increases the
interaction time of light with an acoustic wave that propagates in the
bulk of surrounding water. Furthermore, the tight confinement of the
optical electric field to the metal surface of the nanorod gives 
rise to an increased sensitivity of the plasmon resonance to changes
in the dielectric permittivity caused by the propagation of acoustic
waves occurring in close proximity of the nanorod \cite{Mak15_report}.
At the same time, the light localisation associated with the 
excitation of the higher-order modes results in smaller absorption 
losses compared with the light confinement in small gaps between the
nanodiscs (Fig.~\ref{Fig9}) since the field of higher-order modes does
not penetrate deeply into the metal nanorod.

\begin{figure}[t]
  \centering
  \includegraphics[width=12.0 cm]{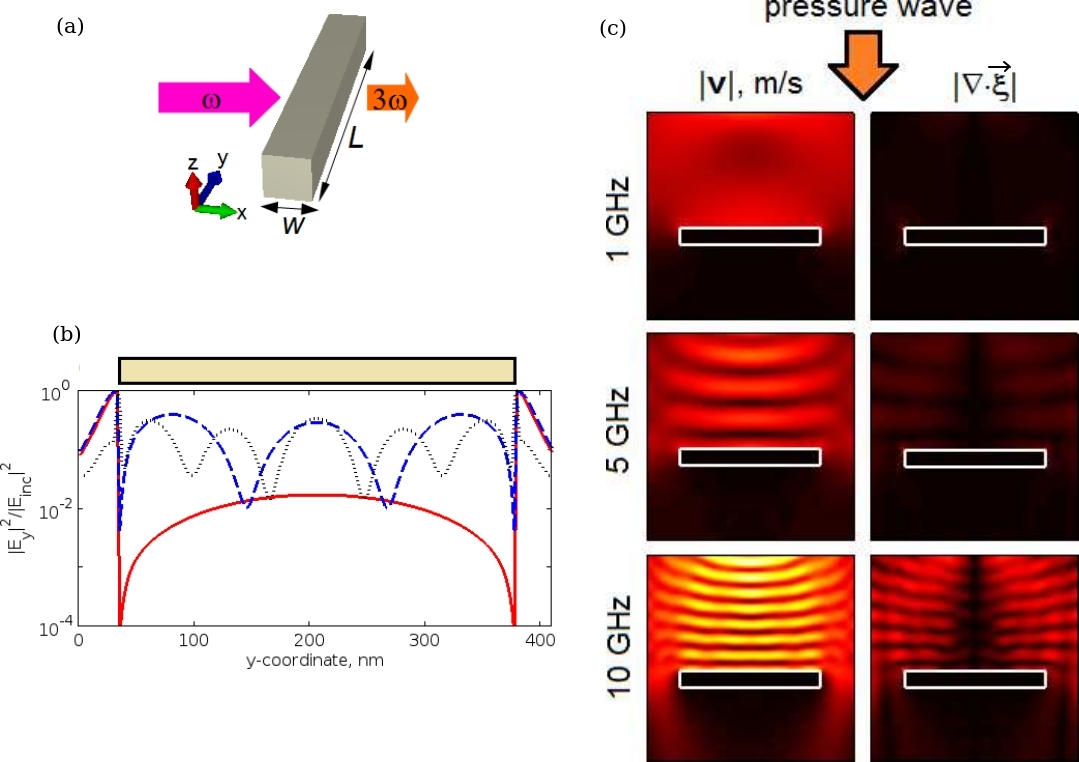}
  \caption{\textbf{(a)}~Illustration of a numerical simulation  
    configuration for investigation of the optical third harmonic 
    generation in a 30\,nm$\times$30\,nm square cross-section silver  
    nanorod immersed into water. A plane light wave polarised along the 
    $y$-axis of the coordinate system is normally incident from the left.  
    \textbf{(b)}~Normalised simulated electric field intensity at the  
    centre of the nanorod cross-section. The rectangle the main panel  
    schematically shows the length of the nanorod. The solid, dashed  
    and dotted lines depict profiles of the fundamental mode, the  
    second and third higher-order modes, respectively.  
    \textbf{(c)}~Simulated spatial profiles of the acoustic velocity  
    |{\bf v}| (left column) and the divergence of the displacement  
    field |$\nabla\cdot\vec\zeta$| for the acoustic frequency 1, 5 and  
    10\,GHz. The white rectangle shows the contour of the  
    nanorod. The black (yellow) colour denotes zero (maximum) of the  
    profile intensity. The direction of the incident monochromatic  
    acoustic pressure wave is indicated by the arrow. Note that the  
    value of |{\bf v}| in front of (behind) the nanorod increases  
    (decreases) due to the well-known pressure doubling effect  
    \cite{Beranek}.\label{Fig10}} 
\end{figure}

There are several approaches to the optical excitation of the 
higher-order plasmonic modes in a long nanorod. Firstly, one can 
tune the frequency of the incident light on the resonance frequency of
a particular higher-order mode. This approach is relatively
straightforward since tuneable laser sources are readily
available. Alternatively, the geometry of a nanorod can be
engineered to match a higher-order mode with a resonance frequency
that coincides with the frequency of the available laser. However,
the drawback of this approach is related to a tight confinement of the
optical fields of the higher-order modes to the surface of a nanorod,
which implies that the energy of such modes is not efficiently emitted
into the electromagnetic far-field region. In turn, this means that
the excitation of these modes by the incident light waves from the
far-field is also insignificant. To overcome this inefficiency the
higher-order modes can be excited using a point-like quantum emitter
of light (e.g.~a quantum dot) located in the vicinity of a
nanorod. However, such an excitation scheme would significantly
increase the complexity of an experiment. An alternative viable
approach could exploit nonlinear-optical properties \cite{Boyd} 
of the nanorod material \cite{Kau12}. In particular, using an intense
laser beam to excite the fundamental mode of a nanorod conditions can
be created for the nonlinear generation of the second and third
harmonics of the incident light \cite{Thy12, Aou13, Pal08}. In this
case, the nanorod length should be so chosen so that the frequency of
one or several of its higher-order modes coincides with the frequency
of the nonlinearity-generated harmonics. The plausibility of this
approach has been confirmed by numerical simulations, where a silver
340\,nm$\times$30\,nm$\times$30\,nm nanorod with a square
cross-section was investigated [Fig.~\ref{Fig10}(a, b)]
\cite{Mak15_report, Mak17}.

The main goal of studies reported in \cite{Mak15_report, Mak17} was
not only to numerically validate an earlier theoretical suggestion
\cite{Mak15_wombat} of a plasmon-enhanced BLS effect in nanorods
immersed in a water or a biological fluid, but also to demonstrate that this
approach could be used to generate an AFC. To achieve such a goal, a
finite-difference time-domain model was developed \cite{Mak15_report}, 
where Maxwell's equations were solved simultaneously with the equations
of nonlinear acoustics. While Maxwell's equations describe the interaction
of light with a metal nanorod and an acoustically-induced fluctuations
of the refractive index of the surrounding liquid, the equations of
nonlinear acoustics describe the interaction of acoustic waves with
the nanorod.

The left column of Fig.~\ref{Fig10}(c) shows the simulated fields
of the sound velocity vector |{\bf v}| in close proximity of a nanorod
for the frequencies of the longitudinal acoustic pressure waves with
frequencies $f_a$=1, 5 and 10\,GHz. The contours of a
340\,nm$\times$30\,nm nanorod are shown by a white rectangle that also
serves as the scalebar. Using these data, the displacement field $\vec\zeta$ of
a particle from its equilibrium position due to the action of acoustic pressure
waves was calculated. Subsequently, the divergence of the displacement
field $\nabla\cdot\vec\zeta$ was computed and shown in the right
column of Fig.~\ref{Fig10}(c). The amplitude of fluctuations of the
dielectric permittivity of water $\delta\epsilon$ caused by the
propagating acoustic pressure wave is directly proportional to
$\nabla\cdot\vec\zeta$ because a plane longitudinal acoustic wave
propagating in the liquid results in alternating compression and
rarefaction and corresponding density changes. The knowledge of the
changes in the dielectric permittivity is required to numerically
simulate the BLS from acoustic waves.

\begin{figure}
  \centering
  \includegraphics[width=12.0 cm]{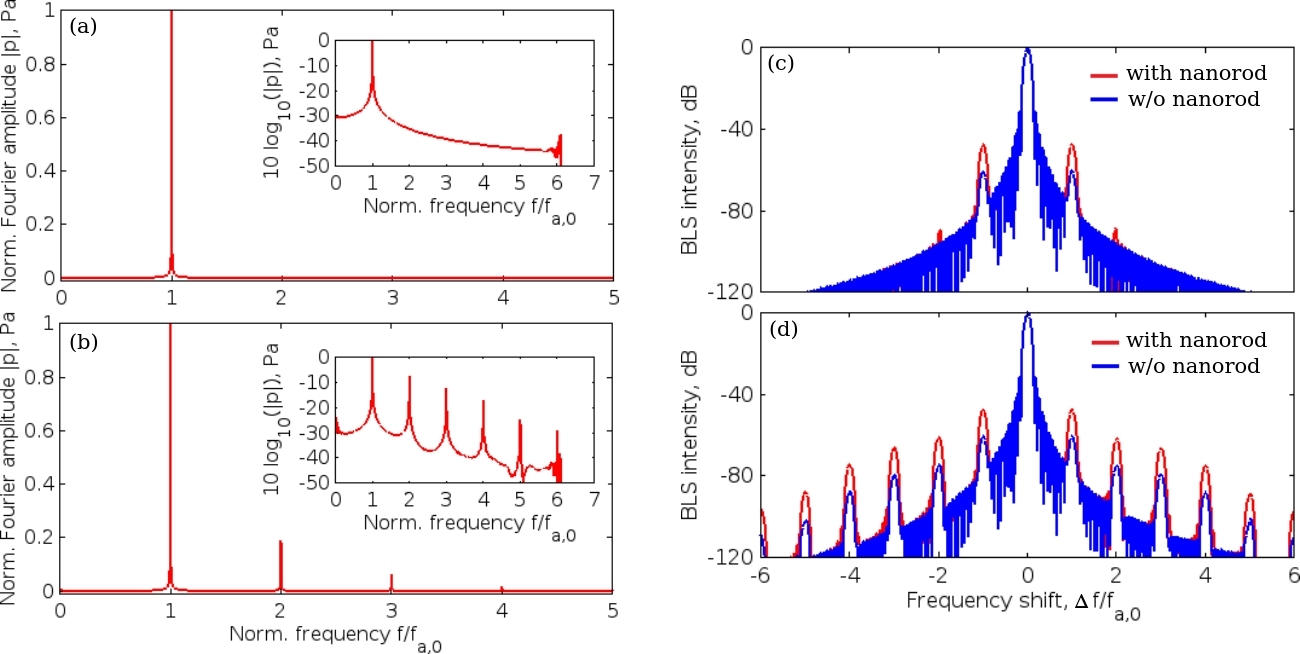}
  \caption{Normalised acoustic energy spectrum of the incident 
    quasi-monochrormatic acoustic pressure wave detected
    \textbf{(a)}~before and \textbf{(b)}~after it propagates through the 
    bulk of water surrounding a nanorod showing the nonlinear 
  generation of acoustic waves at harmonic frequencies of the incident 
  acoustic pressure wave. The frequency is normalised to frequency 
  $f_{a,0}$ of the incident wave. The peak amplitude of the incident 
  wave is 5\,MPa. The insets show the same spectra plotted in a 
  decibel scale. 
  \textbf{(c,~d)}~Simulated plasmon-enhanced intensity BLS signal 
  produced from acoustic signals with the spectra shown in Panels~(a) 
  and (b). Note an approximately 35-fold enhancement due to plasmonic 
  properties of a nanorod.\label{Fig11}}
\end{figure}       
The simulated spectrum of the incident acoustic wave is shown in 
Fig.~\ref{Fig11}(a) and its inset, where the same spectrum plotted
in the decibel scale. Only one peak at the normalised frequency
$f/f_{a,0}=1$ can be seen confirming the monochromatic nature of this
wave (since qualitatively the same result was obtained for kHz, MHz
and GHz frequency range waves, the frequency was normalised to the  
frequency $f_{a,0}$ of the incident wave). However, when acoustic
nonlinearity develops as a result of acoustic wave propagation in
water, additional normalised frequency peaks appear in the spectrum at
$f/f_{a,0}=2$ (quadratic nonlinearity effect), $f/f_{a,0}=3$ (cubic
nonlinearity effect) and so on, see Fig.~\ref{Fig11}(b)]. In addition
to the appearance of the wave harmonics, the nonlinear interaction
leads to the creation of a mean drift (i.e.~zero frequency)
component that can be seen on logarithmic scale.

The time-domain dependencies of the dielectric permittivity
fluctuations $\delta\epsilon(t)$ were also extracted from the 
numerical data and used in simulations of the plasmon-enhanced
BLS effect, the results of which are presented in Fig.~\ref{Fig11}(c, d).
Using the simulation data reported in Fig.~\ref{Fig11}(a) a typical
BLS spectrum with the central Rayleigh peak and two weak side peaks
shifted by the normalised frequency of the incident acoustic wave 
$\Delta f/f_{a,0}=\pm1$ from it was found, see
Fig.~\ref{Fig11}(c). Note that the amplitude of the shifted Brillouin
peaks is significantly increased when a nanorod is present due to the
plasmonic enhancement. The presence of a nanorod also leads to the
appearance of the second order Brillouin peaks shifted by $\Delta
f/f_{a,0}=\pm2$ from the central peak. Most importantly, due to  a
strong acoustic nonlinearity present in Fig.~\ref{Fig11}(d) the
generation of an AFC is observed with the spectrum consisting of
Brillouin peaks shifted by  
$\Delta f/f_{a,0}=\pm1,\pm2,\pm3$ and so on with
respect the to the central peak. In the presence of a nanorod the
amplitude of all these peaks is increased due to its plasmonic
behaviour. It was also demonstrated in the follow-up paper
\cite{Mak17}, where a more advanced model of the plasmon-enhanced BLS
interaction was proposed, that all plasmon-enhanced Brillouin peaks
are phase coherent and thus can be considered as a FC similar to a
mode-locked laser OFC \cite{Pic19}. Finally, we note that
qualitatively similar results were obtained for the values of
$f_{a,0}$ in a wide range from several kHz to several GHz, which
implies that the inter-peak distance of the AFCs discussed in this
section can also lie in this wide frequency range. 

\section{Frequency comb generation using oscillations of gas bubbles
  in liquids} 
\subsection{Physical origin of the acoustic nonlinearity of gas
  bubbles} 
From the discussion preceding this section, it becomes clear that
strong acoustic nonlinearities can result in the generation 
of AFCs containing a large number of high-amplitude peaks.
Similar conclusions can be drawn regarding the Kerr OFCs and other
FC generation techniques relying on nonlinear optical
phenomena. However, in the field of Kerr OFCs there exist
  fundamental physical limitations that do not allow one to increase
  the intensity of the laser beam indefinitely to amplify nonlinear
effects and thus to increase the number of peaks in the spectrum of the
comb \cite{Mak19}. While special techniques and novel materials have
been proposed to relax such limitations \cite{Sal08, Fer08, Li12, Ala16}
either by optimising the conversion of the energy of the incident light
into new frequency signals or minimising optical absorption losses,
in general they cannot be removed completely.

However, recently it has been theoretically demonstrated that
nonlinear optical effects could be effectively replaced by acoustic
nonlinearities. For example, a change in the refractive index due to a
propagating acoustic pressure wave can modulate an optical signal 
and, if the acoustic wave exhibits nonlinearities, these also become
imprinted onto the modulated optical signal effectively enabling the
conversion of the acoustic nonlinearity into new optical signals
\cite{Mak19}. Such a nonlinear acousto-optical interaction may
generate additional optical frequencies.

This concept of a hybrid nonlinear acousto-optical interaction
exploits the fact that acoustic nonlinearities are much stronger than
their optical counterparts and that they can be induced using sound
waves with a relatively low peak pressure amplitude (recall that the
field of nonlinear optics was established only after the invention of
powerful lasers since required to induce nonlinear optical effects). 
In this context, it is worth noting the so-called giant acoustic 
nonlinearities associated with the oscillations of gas bubbles in
liquids \cite{Ray17, Min33, Ple49, Pro74, Fra83, Fra84, Kel80, Bre95,
  Met97, Rud06, Lau10, Doinikov_book, Sus12, Dza13}. When an
  acoustic pressure wave propagates through water, its initially
  sinusoidal waveform changes so that its initial monochromatic
  spectrum acquires higher harmonic frequencies, see
  Fig.~\ref{Fig11}(a) and (b). The more nonlinear the medium in which
  sound propagates is, the stronger such a spectral enrichment. The
  degree of acoustic nonlinearity is often characterised by the
  acoustic parameter $\beta=B/A$, which is the ratio of coefficients
  $B$ and $A$ of quadratic and linear terms in the Taylor series
  expansion of the equation of state
  \[p=p(\rho)\approx p_0+A\dfrac{\rho-\rho_0}{\rho_0}
  +B\dfrac{(\rho-\rho_0)^2}{2\rho_0^2}+\cdots\]
of a medium relating the thermodynamic pressure $p$ in the medium with
its density $\rho$, where subscript 0 denotes the values in the
absence of sound \cite{Rud06, Mak19}. The larger the value of $\beta$
is, the more nonlinear the medium, the stronger a distortion of the
acoustic spectrum from the initial monochromatic state. For example,
water with $\beta=3.5$ is more acoustically nonlinear than air with
$\beta\approx0.7$ \cite{Mak19}, but the degree of nonlinearity is
moderate in both media. However, when air bubbles are injected in water,
the value of $\beta$ increases to around 5000 \cite{Rud06, Mak19}. The
following qualitative discussion explains this fact. 

Liquids are dense and have little free space between molecules, which
leads to their low compressibility. On the contrary, gases are easily
compressible. When an acoustic wave propagating in water reaches a 
bubble, due to high compressibility of air trapped in it its volume
changes dramatically (see, e.g., \cite{Lau10}). This in turn results
in large local acoustic wavefront deformations that result in strong
variation of the initial acoustic spectrum. Consequently, whereas in
bubble-free water one can observe the generation of five or so higher
frequency harmonics of the incident sound as seen from
Fig.~\ref{Fig11}(b), in bubbly water up to 20 high-order acoustic
harmonics can be generated for the acoustic wave with the same peak
pressure amplitude effectively forming AFC.
\begin{figure}[t]
  \centering 
  \includegraphics[width=12.0 cm]{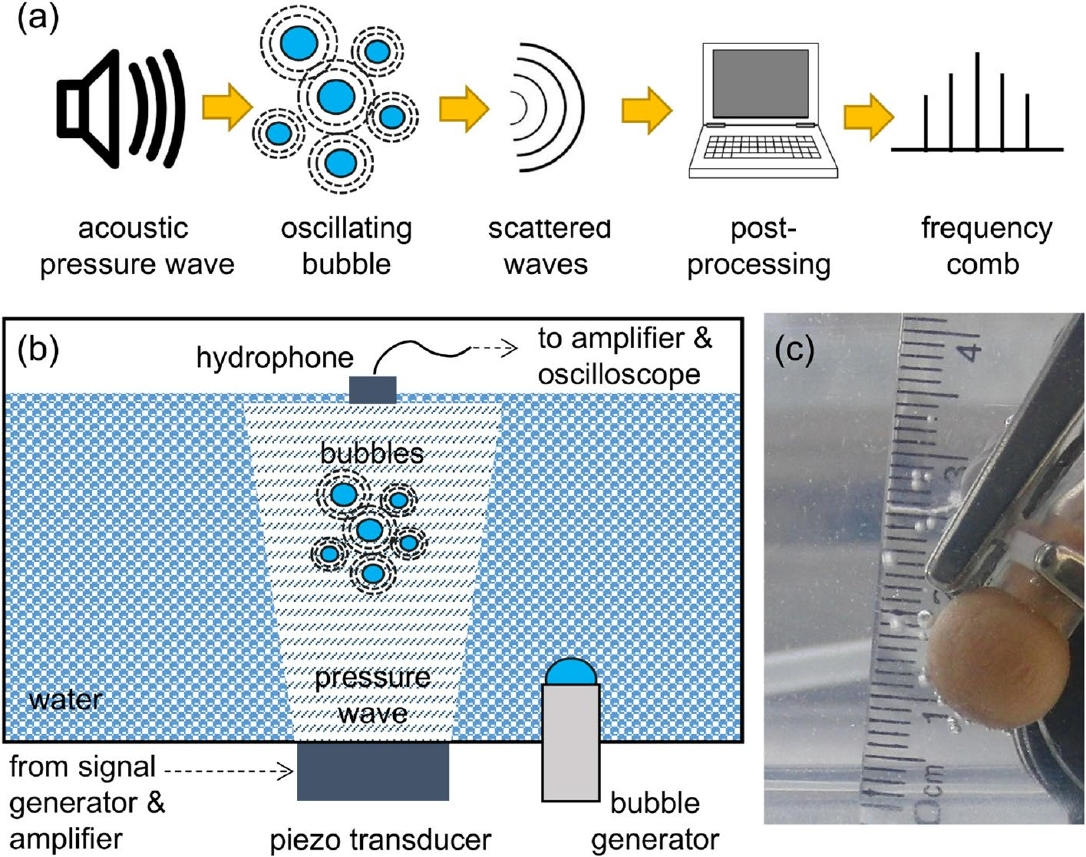}
  \caption{\textbf{(a)}~Schematic diagram of the AFC generation 
    using oscillations of gas bubbles in water. The oscillations are 
    driven by a single-frequency ultrasound pressure wave. Acoustic 
    waves scattered by the bubbles are recorded and post-processed to 
    obtain a spectrum consisting of equidistant peaks. 
    \textbf{(b)}~Schematic of an experimental setup, where bubbles are 
    created in a stainless steel tank using a bubble generator, the 
    driving pressure wave is emitted by an ultrasonic transducer, and 
    waves scattered by the bubbles are detected by a hydrophone. 
    \textbf{(c)}~Photograph of typical gas bubbles emitted by a bubble 
    generator. Reproduced from \cite{Mak21} published by Springer Nature
    under the terms of the Creative Commons CC BY license. 
    \label{Fig12}} 
\end{figure} 

\subsection{Acoustic frequency comb generation using oscillations of
multiple gas bubbles in water}
\begin{figure}
  \centering
  \includegraphics[width=12.0 cm]{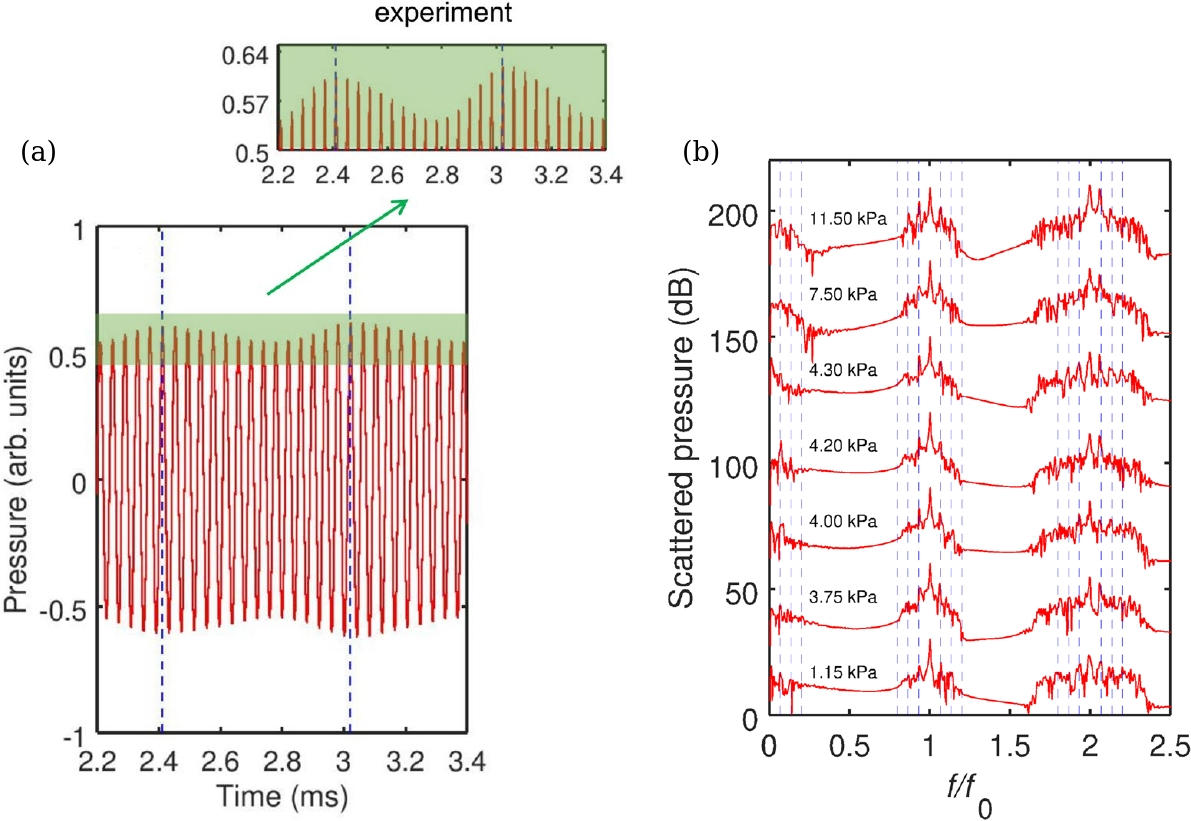}
  \caption{\textbf{(a)}~Measured time-domain acoustic response of 
    gas bubbles. The time between the vertical dashed lines is 
    $\Delta T=1/f_{nat}\approx0.6$\,ms, where $f_{nat}$ is the natural 
    frequency of the bubble oscillations (see \cite{Mak21} for details). 
    The insets show the closeup of the waveforms and demonstrate the 
    amplitude modulation.
    \textbf{(b)}~Experimental AFC spectra obtained using gas bubbles
    in water insonated with an $f_0=24.6$\,kHz sinusoidal signal of
    increasing pressure amplitude $\alpha$=1.15, 3.75, 4, 4.2, 4.3,
    7.5 and 11.5\,kPa. The scattered pressure values (in dB) are shown
    along the vertical axis with the vertical offset of 30\,dB
    between the spectra. Reproduced from \cite{Mak21} published by
    Springer Nature under the terms of the Creative Commons CC BY license. 
    \label{Fig13}}
\end{figure}

The ideas discussed above have been validated experimentally in
\cite{Mak21}, where a single-frequency ($f_0=24.6$\,kHz) ultrasound
wave irradiated several gas bubbles created in a water tank using a
gas bubble generator, see Fig.~\ref{Fig12}. A small (not exceeding
11.5\,kPa) peak pressure amplitude of the driving ultrasound wave was
deliberately chosen since, as discussed above, low-amplitude signals
suffice to induce strong nonlinearities in liquid-gas mixtures. The
generated bubbles had the equilibrium radii $R_0\approx 1.0\pm
0.5$\,mm. However, since they interacted with each other during the
oscillation driven by the ultrasound wave, a collective acoustic
response typical of a small bubble cluster with an effective natural
frequency \cite{Min33, Lau10} $f_{nat}\approx1.7$\,kHz was
observed. Using high-speed imaging and following \cite{Hwa00}, it was
estimated that the resulting cluster behaved similarly to a large
single gas bubble with an equilibrium radius of 1.95\,mm. Thus, since
$f_{nat}$ is an order of magnitude lower than the frequency of the
ultrasound wave, the oscillations of the cluster of bubbles resulted
in a nonlinear generation of multiple ultraharmonic frequency peaks in
the spectrum of the cluster's acoustic response. The interaction of
the so-generated acoustic waves with the noise-induced bubble
oscillations at the natural frequency resulted in the amplitude
modulation of the collective bubble response [Fig.~\ref{Fig13}(a)],
and the appearance of sidebands around the harmonic and ultraharmonic
peaks [Fig.~\ref{Fig13}(b)]. These sideband structures can be used as AFCs.

\begin{figure}
  \centering
  \includegraphics[width=12.0 cm]{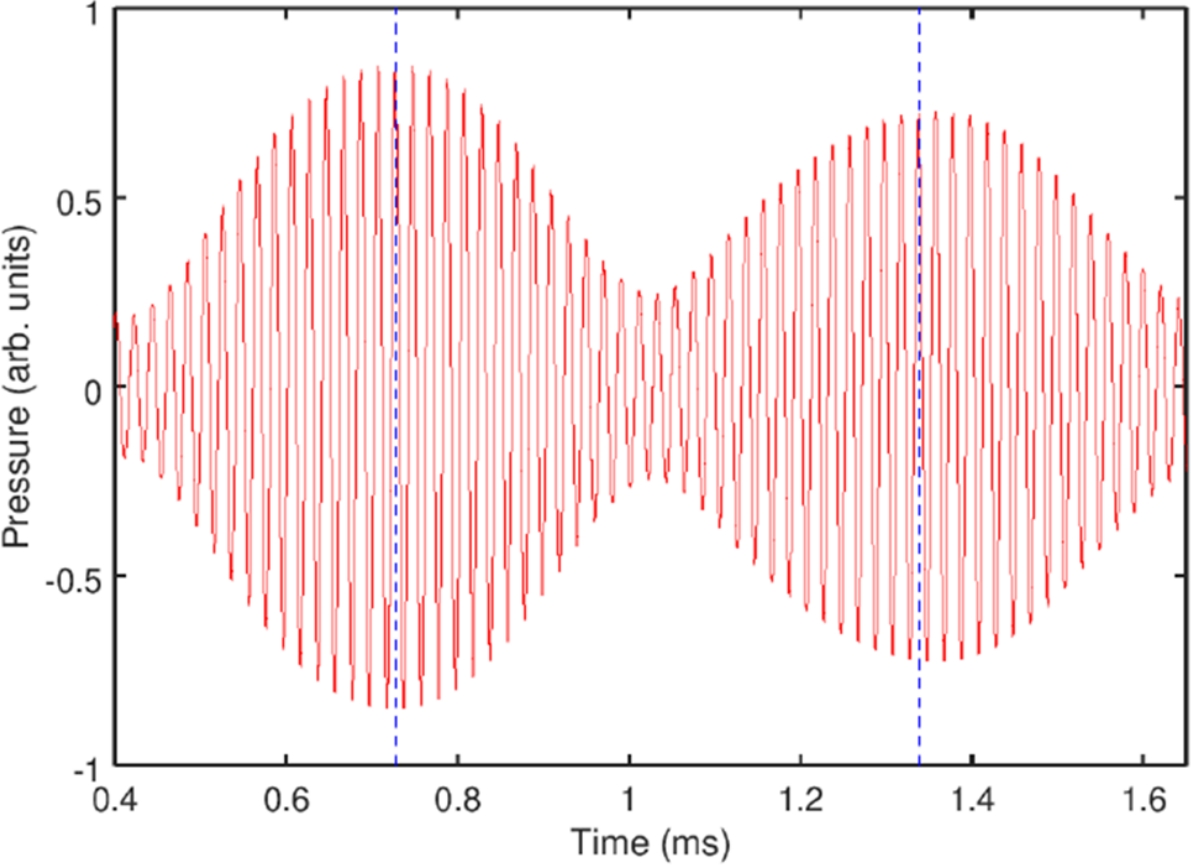}
  \caption{Measured acoustic bubble response corresponding to a 
    sinusoidal driving pressure wave with the frequency 
    $f_0=49.2$\,kHz (twice the frequency in Fig.~\ref{Fig13}(a)) and 
    amplitude $\alpha = 4.3$\,kPa. Reproduced from \cite{Mak21}
    published by Springer Nature under the terms of the Creative
    Commons CC BY license.\label{Fig14}}
\end{figure}

Several comments should be made to better explain this
result. Firstly, in the temporal profile in Fig.~\ref{Fig13}(a), the
amplitude modulation depth (the ratio of the modulation excursions to
the amplitude of the unmodulated carrier wave) is smaller than 1. In
some OFC technologies, where direct photodetection of the optical
pulses is used to produce an electronic signal that follows the
amplitude modulation of the pulse train, low modulation depth could
pose technical challenges. However, this does not present a problem in
the case of AFC using oscillating gas bubbles since the frequencies of
the electronic signal of the AFC coincide with the frequencies of the
driving pressure wave, which dramatically simplifies the
characterisation of the comb. Secondly, in \cite{Mak21} the analysis
of the experimental result was supported by a rigorous theoretical and
computational modelling of gas bubbles oscillations, and it was shown
that both experimental time-domain signals and AFC spectra are in good
qualitative agreement with the calculated ones. However, because the
numerical model in \cite{Mak21} considered only a single oscillating
gas bubble with an effective equilibrium radius of 1.95\,mm, it was
unable to reproduce some experimentally observed features, namely, the
generation of another AFC spectrum centred at the second harmonic of
the driving ultrasound wave. In fact, Fig.~\ref{Fig13}(b) shows a
sideband peak structure at 49.2\,kHz (i.e.~$f/f_0 = 2$) and peak
ultrasound wave amplitude $\alpha=4.3$\,kPa. As shown in
Fig.~\ref{Fig14}, at this frequency the amplitude modulation also
gives rise to a train of pulses with the modulation period close to
that of the natural bubble cluster oscillations confirming that this
signal can also be used as an AFC. Thirdly, a slight irregularity of
the AFC peaks in Fig.~\ref{Fig13}(b) was attributed to the Doppler
effect associated with a translational motion of oscillating bubbles
in the incident ultrasound field \cite{Doinikov_book}. The size
variation of the generated bubbles could also contribute to the comb
peak imperfection. However, these deficiencies were not considered as
prohibitive, which is demonstrated in the following section.

\subsection{Spectrally-wide acoustic frequency combs generated using
oscillations of polydisperse gas bubble clusters in liquids}
As with the other AFC generation techniques \cite{Gan17, Gar18}
discussed in this review article, in the experiments of \cite{Mak21}
the number of the sideband peaks usable as an AFC is
limited. Currently, this presents numerous technological challenges
that shape research efforts in the field of AFCs (similar problems
also exist in the field of OFCs \cite{Pic19, For19}). For example, for
many applications the spectrum of an AFC has to span over an octave of
bandwidth (i.e.~the highest frequency in the FC spectrum has to be at
least twice the lowest frequency). To achieve this, the spectrum of an
AFC can be extended using one of the techniques developed, for
example, for broadening the spectra of opto-electronic FCs
\cite{Zha19_1} such as supercontinuum generation using nonlinear
optical effects (as demonstrated above, the adoption of optical
techniques in acoustics is possible because of the analogy between
nonlinear optical processes in photonic devices and nonlinear acoustic
processes in liquids containing gas bubbles). Furthermore, the
theoretical analysis in \cite{Mak21} demonstrated that the number of
peaks in a nonlinearly generated AFC and their magnitude can
be increased by simultaneously decreasing the frequency and increasing 
the pressure of the ultrasound wave driving bubble oscillations. 

As seen from Fig.~\ref{Fig13}(b), the shape of spectral peaks is
slightly irregular and it could be argued that this artefact is
associated with a translational motion of oscillating bubbles in the
incident ultrasound field. Therefore, the question of long-term
stability of AFC signals arises, which was comprehensively addressed
in \cite{Tony21}.

The interplay between radial bubble oscillations and their
translational motion has been a subject of intensive research
\cite{Nem83, Zab84, Wat93, Pel93, Doi95, Met97, Bar99, Har01, Doi01,
  Mat05, Mac06, Met09, Yas10, Sad10, Jia12, Dza13, Lan15, Leb15,
  Doinikov_book}. Most of these studies are based on the accepted
models of spherical gas bubble oscillations \cite{Ray17, Ple49, Kel80} 
and consider Bjerknes forces \cite{Bje06} acting on oscillating
bubbles. The primary Bjerknes force $F_{pB}$ is caused by the acoustic
pressure field forcing a bubble \cite{Bje06, Lei90} while the
secondary Bjerknes force $F_{sB}$ arises between two or more
interacting bubbles \cite{Doinikov_book}. The secondary Bjerknes force
between two gas bubbles is repulsive when the driving frequency lies
between bubbles' natural frequencies, otherwise it is attractive
\cite{Kaz60, Cru75, Doinikov_book}.

In \cite{Tony21}, an alternative strategy for broadening spectra of
AFCs generated using gas bubble oscillations was suggested using
polydisperse clusters consisting of mm-sized bubbles with equilibrium 
radii $R_{n0}=R_{10}/n$, where $R_{10}$ is the equilibrium radius of the
largest bubble in the cluster and $n=1,2,3,\dots$ is the total number
of bubbles. Although clusters with other bubble size distributions 
could also be used in the proposed approach, it was shown that this specific
ratio of equilibrium radii enables generating AFCs with a quasi-continuum
of equally spaced peaks. Similarly to the experiment \cite{Mak21}, in the
analysis in \cite{Tony21} low-pressure ultrasound waves (up to 10\,kPa)
were considered and a numerical model of dynamics of multibubble clusters
with translational motion developed in \cite{Doi01, Doi04, Doinikov_book}
was employed. 

\begin{figure}[t]
  \centering
  \includegraphics[width=12.0 cm]{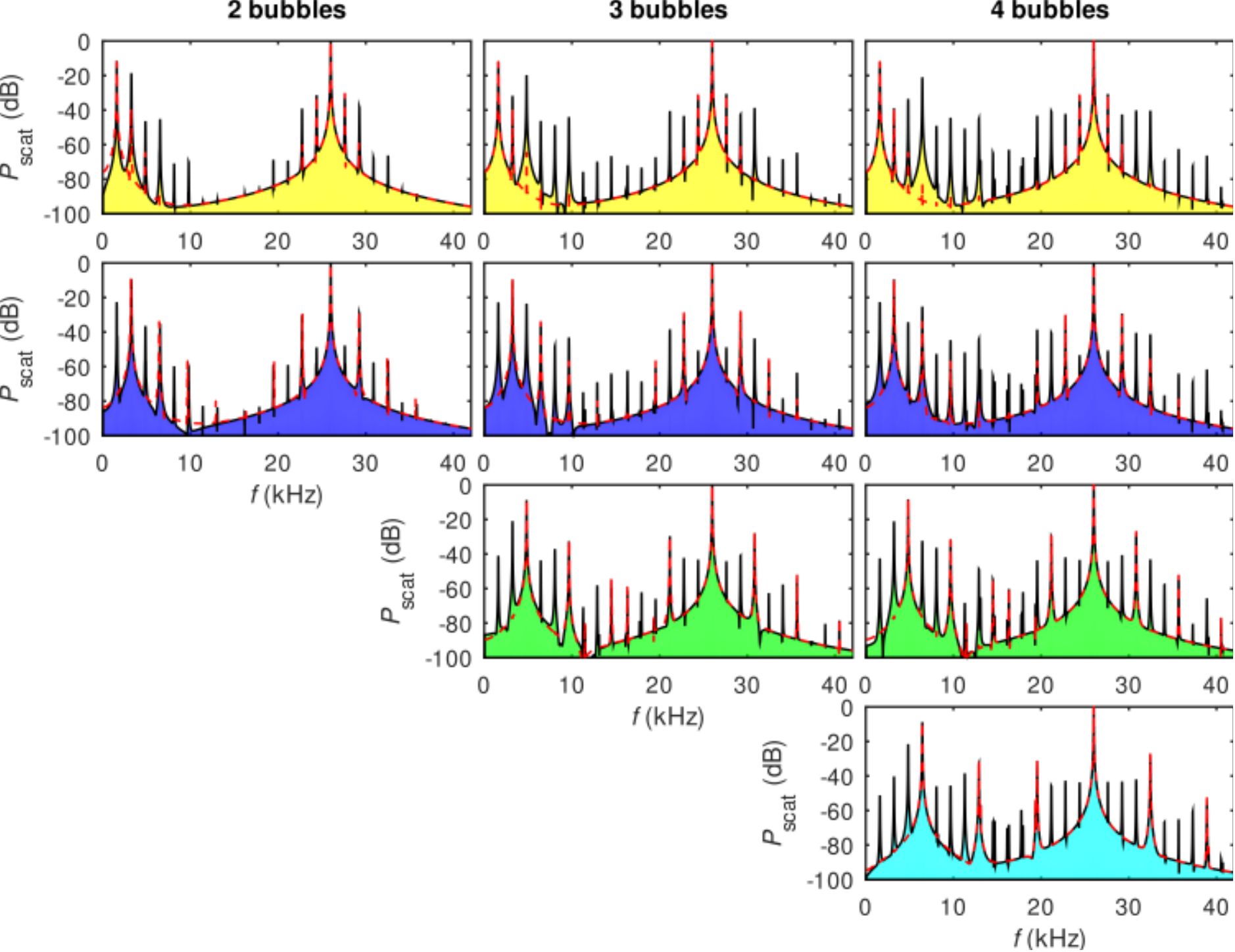}
  \caption{Columns (from left to right) show the AFC spectra produced
    by individual bubbles within clusters consisting of two, three,
    and four bubbles with the equilibrium radii $R_{n0}=1.95/n$\,mm,
    where $n$ is the bubble index in the cluster. The number of panels
    in each column corresponds to the total number of bubbles. The red  
    dashed lines in each panel show the spectra of individual
    non-interacting stationary bubbles with identical equilibrium
    radii. Computational parameters are given in
    \cite{Tony21}. Reproduced from \cite{Tony21}. Copyright 2021 by the
    American Physical Society.\label{Fig15}}
\end{figure}

Figure~\ref{Fig15}(a) shows the calculated spectra of the bubble
clusters, where the number of rows in each column corresponds to the
total number of bubbles in the cluster. Each column shows the spectrum
of the pressure scattered by an individual bubble within the
cluster. The inspection of panels within the same row from left to
right reveals changes in the AFC peak structure caused by the addition
of smaller bubbles to the cluster. For example, the four panels in the
top row show that the number of equidistant peaks in the AFC spectrum
produced by the largest bubble increases when smaller bubbles are
added. This is because bubbles within a cluster are affected by the
pressure waves scattered by their neighbours and thus their spectra
include additional frequency peaks compared to the spectra of isolated
noninteracting stationary bubbles of the same equilibrium radii (the
dashed lines in Fig.~\ref{Fig15}). Similarly, panels in the second row
show the evolution of the AFC spectrum of the second largest bubble in
the cluster, and so on. In all cases the spectra exhibit a key
features of a pure AFC signal---the spectrum of the acoustic response
of each bubble consists of a series of well-defined equally spaced
peaks.

\subsection{Temporal stability of bubble-based acoustic frequency
  combs} 
Numerical modelling reported in \cite{Tony21} demonstrated that the
frequency of an ultrasound wave driving bubble oscillations can be
chosen in a wide spectral range above the natural oscillation
frequency of bubbles in a cluster. This can greatly facilitate the 
generation and recording of stable AFC signals since at low pressure
bubble clusters exhibit a regular behaviour for a longer time before
their dynamics becomes affected by bubble aggregation. To examine the
basic trends in the temporal stability of a bubble cluster, here we
consider a system of just two interacting bubbles in liquid. This
simplification allows reducing the complexity of the model while still
accounting for essential physics of bubble interaction.

The accepted model of nonlinear oscillations of a single spherical gas
bubble that does not undergo translational motion is the Keller-Miksis 
equation \cite{Kel80}. It takes into account the decay of bubble
oscillations due to viscous dissipation and fluid
compressibility. However, for mm-sized gas bubbles oscillating at
20--100\,kHz frequencies in water being driven by low-pressure
ultrasound waves with the amplitude of up to 10\,kPa the terms in the
KM equation accounting for acoustic losses become negligible
\cite{Tony21}. Thus, the Keller-Miksis equation effectively reduces to the
classical Rayleigh-Plesset equation \cite{Ray17, Ple49}, which is
written for a cluster consisting of $N$ mm-sized gas bubbles not
undergoing translational motion and being driven by low-pressure
ultrasound waves as \cite{Met97, Dza13} 
\begin{equation} 
  R_n\frac{d^2R_n}{dt^2}+\frac{3}{2}\left(\frac{dR_n}{dt}\right)^2
  =\frac{1}{\rho}\left(P\left(R_n,\frac{dR_n}{dt}\right)
    -P_\infty(t)\right)-P_{sn}\,,\label{eq:eq5}
\end{equation}
where
\begin{equation}
  P_n\left(R_n,\frac{dR_n}{dt}\right)=\left(P_0-P_v+\frac{2\sigma}{R_{n0}}\right)
  \left(\frac{R_{n0}}{R_n}\right)^{3\kappa} 
  -\frac{4\mu}{R_n}\frac{dR_n}{dt}-\frac{2\sigma}{R_n}\,.\label{eq:eq6}
\end{equation}
The term accounting for the pressure acting on the $n$th bubble due to
scattering of the incoming pressure wave by the neighbouring bubbles
in a cluster is given by 
\begin{equation}
  P_{sn}=\sum_{l=1, l\neq n}^N\dfrac{1}{d_{nl}}\left( R_l^2\frac{d^2R_l}{dt^2}
    + 2R_l\left( \frac{dR_l}{dt}\right)^2 \right)\,,\label{eq:eq6_1}
\end{equation}
where $d_{nl}$ is the inter-bubble distance \cite{Met97}. The pressure
in a liquid far from the bubble is represented by
$P_\infty(t)=P_0-P_v+\alpha\sin(\omega^*t)$ with the angular frequency  
$\omega^*=2\pi f$, where $P_0$ and $P_v$ are the air and vapor
pressures, respectively. Parameters $R_{n0}$, $R_n(t)$, $\mu$, 
$\rho$, $\kappa$, $\sigma$, $\alpha$ and $f$ denote the equilibrium
and instantaneous radii of the $n$th bubble in the cluster, the
dynamic viscosity and the density of the liquid, the polytropic
exponent of a  gas entrapped in the bubble, the surface tension of a
gas-liquid interface and the amplitude and the frequency of a driving
ultrasound wave. Diffusion of the gas through the bubble surface is
neglected.

To identify the main characteristics of nonlinear oscillations of
interacting gas bubbles relevant to the generation of AFCs, an
asymptotic analysis of Eq.~\eqref{eq:eq5} is conducted after it is
rewritten in the non-dimensional form using the equilibrium radius of
the largest bubble in the cluster, $R_{10}$, and $1/{\omega^*}$ as the
length and time scales, respectively, to introduce the non-dimensional
quantities 
$r_n={R_n(t)}/R_{10}$, $r_l={R_l(t)}/R_{10}$ and $\tau^* = \omega^*t$
\cite{Dza13}. This results in
\begin{eqnarray}
  r_n {r_n}''
  +\frac{3}{2} {{r_n}'}^2
  &=&\left(\M+\frac{\W}{\Q_n}\right)\left(\frac{\Q_n}{r_n}\right)^\K
      -\frac{\W}{r_n}-\R\frac{{r_n}'}{r_n}-\M-\M_e\sin\tau^*\nonumber\\
  &&-\sum_{\substack{l=1\\l\neq n}}^N\zeta_{nl}
  \left(r_l^2r_l''+2r_l{r_l'}^2\right)\,,\label{eq:eq6_2}
\end{eqnarray}
where $\R=\dfrac{4\mu}{\rho\omega^*R^2_{10}}$,
$\W=\dfrac{2\sigma}{\rho\omega^{*2}R^3_{10}}$,
$\M=\dfrac{P_0-P_v}{\rho\omega^{*2}R^2_{10}}$,
$\M_e=\dfrac{\alpha}{\rho\omega^{*2}R^2_{10}}$,
$\zeta_{nl}=\dfrac{R_{10}}{d_{nl}}$,
$\K=3\kappa$ and $\Q_n=\dfrac{R_{n0}}{R_{10}}$.
Parameters $\R$ and $\W$ can be treated as inverse Reynolds and Weber
numbers representing the viscous dissipation and surface tension
effects, respectively. Parameter $\M$ characterises the ratio of
bubble's natural and forced oscillation frequencies and $\M_e$ is the
measure of the ultrasound forcing \cite{Sus12}. Parameters
$\zeta_{nl}$ and $\Q_n$ are the inverse of the distance between the
bubble centres and the bubble radius relative to that of the largest
bubble in the cluster, respectively \cite{Dza13} and primes denote
differentiation with respect to $t$. As discussed in \cite{Sus12,
  Mak21, Tony21}, $\K=4$ for air and for bubbles of sizes
relevant to the AFC context the maximum values of other parameters do
not exceed $\M=9.7\times10^{-4}$, $\W=7.4\times10^{-7}$, 
$\R=6.5\times10^{-6}$ and $\M_e=9.9\times10^{-5}$. Therefore, the
effects of water viscosity and surface tension on bubble oscillations
are negligible and we set $\R=\W=0$ in what follows. Thus,
ultrasonically forced bubble oscillations can be assumed perfectly
periodic when the driving frequency is much higher than any
of the natural frequencies of the individual bubbles in the
cluster (i.e.~no resonances arise). This warrants using analysis
similar to that of \cite{Mak21}.

Consider a cluster consisting of two gas bubbles with the
non-dimensional equilibrium radii $r_{n0}=\Q_n$, $n=1,2$
($\Q_1\equiv1$). Following \cite{Che07, Mak21} we look for the
asymptotic solutions of Eq.~\eqref{eq:eq6_2} in the form  
\begin{equation}
  r_n=\Q_n+\epsilon r_{n1}(\tau)+\epsilon^2r_{n2}(\tau)
  +\ldots\,,\quad n=1,2\,,\label{eq631}
\end{equation}
where $0<\epsilon\ll1$ is a parameter characterising the amplitude of
bubble oscillations used to distinguish between various terms in the
asymptotic series, $\tau=\omega\tau^*=\omega\omega^*t$ and
$\omega=\sqrt{\K\M}$ is Minnaert frequency \cite{Min33, Mak21}
of the largest bubble in the cluster. At the first order of $\epsilon$
we obtain
%\begin{eqnarray}
%  \ddot r_{11}+\frac{\K\M}{\Q_1^2\omega^2}r_{11}
% +\frac{\Q_2^2}{\Q_1}\zeta_{12}\ddot{r}_{21}
%  &=&\frac{p}{\Q_1}\sin(\Omega\tau)\,,\\
%  \ddot r_{21}+\frac{\K\M}{\Q_2^2\omega^2}r_{21}
%  +\frac{\Q_1^2}{\Q_2}\zeta_{12}\ddot{r}_{11}
%  &=&\frac{p}{\Q_2}\sin(\Omega\tau)\,,  
%\end{eqnarray} 
\begin{eqnarray}
  \ddot{r}_{11}+r_{11}+\Q_2^2\zeta_{12}\ddot{r}_{21}
     &=&p\sin(\Omega\tau)\,,\label{eqr11}\\
  \ddot{r}_{21}+\frac{1}{\Q_2^2}r_{21}+\frac{\zeta_{12}}{\Q_2}\ddot{r}_{11}
     &=&\frac{p}{\Q_2}\sin(\Omega\tau)\,.\label{eqr21}
\end{eqnarray}
where overdots denote differentiation with respect to $\tau$,
$(\M_e/\omega^2)\sin{\tau^*}\equiv-\epsilon p\sin(\Omega\tau)$  
and $\Omega\equiv1/\omega\gg1$. At $\cO(\epsilon^2)$ equations become:
\begin{eqnarray}
  &\ddot{r}_{12}+r_{12}
  +\Q_2^2\zeta_{12}\ddot{r}_{22}
  =\dfrac{\K+1}{2}r_{11}^2
      -\dfrac{3}{2}\dot{r}_{11}^2-r_{11}\ddot{r}_{11}
      -2\Q_2\zeta_{12}\left(\dot r_{21}^2+r_{21}\ddot r_{21}\right)\,,&
      \label{eqr12}\\
  &\ddot r_{22}+\dfrac{1}{\Q_2^2}r_{22}
  +\dfrac{\zeta_{12}}{\Q_2}\ddot r_{12}
  =\dfrac{\K+1}{\Q_2^2}r_{21}^2
      -\dfrac{3}{2\Q_2}\dot r_{21}^2-\dfrac{1}{\Q_2}{r}_{21}\ddot r_{21}
      -2\dfrac{\zeta_{12}}{\Q_2}\left(\dot{r}_{11}^2+r_{11}\ddot r_{11}\right)
      \,.&\label{eqr22}
\end{eqnarray}
Following \cite{Mak21} we write the random initial conditions as
$r_{1}(0)=1+\epsilon a$, $r_{2}(0)=\Q_2+\epsilon b$, 
$\dot{r_1}(0)=\epsilon c$ and $\dot{r_2}(0)=\epsilon d$ that results
in  
\begin{equation}
   r_{11}(0)=a\,,\quad  r_{21}(0) =b\,,\quad \dot{r}_{11}(0)=c\,,\quad
   \dot{r}_{21}(0)=d\,.
    \label{eq:randIC}
\end{equation}
Subsequently, we obtain the leading order solutions
\begin{eqnarray}
  r_{11}(\tau)
  &=&B_1\sin{\Omega\tau}
      +C_{11}\cos(\omega'_1\tau)+C_{12}\sin(\omega'_1\tau)\nonumber\\	
  &&+C_{21}\cos(\omega'_2\tau)+C_{22}\sin(\omega'_2\tau)\,,\label{eq:1order1}\\
  r_{21}(\tau)
  &=&B_2\sin{\Omega\tau}
      +\dfrac{1-{\omega_1'}^2}{{\omega_1'}^2\Q_2^2\zeta_{12}}
      (C_{11}\cos(\omega'_1\tau)+C_{12}\sin(\omega'_1\tau))\nonumber\\
  &&+\dfrac{1-{\omega_2'}^2}{{\omega_2'}^2\Q_2^2\zeta_{12}}
     (C_{21}\cos(\omega'_2\tau)+C_{22}\sin(\omega'_2\tau))\,,\label{eq:2order1}\\
  \omega'_{1,2}&=&\sqrt{\dfrac{2}{\Q_2^2+1
		\pm\sqrt{(\Q_2^2-1)^2+4\Q_2^3\zeta_{12}^2}}}\,.
\end{eqnarray} 
These frequencies depend on the inverse inter-bubble distance
$\zeta_{12}$ \cite{Zab84,Doinikov_book, Man16}. Considering a
particular case of $\Q_2=\frac{1}{2}$, as expected, for
non-interacting distant bubbles with $\zeta_{12}\rightarrow0$ we
obtain $\omega'_1\to\omega'_{10}=1$ and $\omega'_2\to\omega'_{20}=2$.
In general, the leading order bubble response will always contain
three distinct frequencies: two bubble’s natural frequencies
$\omega'_{1,2}$ and the driving ultrasound frequency $\Omega$.

Coefficients $B_i$ and $C_{ij}$, $i,j=1,2$
% and phase shifts $\phi_{1,2}$
in Eqs~\eqref{eq:1order1} and \eqref{eq:2order1} depend on $\zeta_{12}$, 
$\Omega$ and $p$ and can be obtained for arbitrary initial conditions
(\ref{eq:randIC}). However, their expressions are too long to be given
here explicitly. We only state that they demonstrate that the
magnitude of the $\omega_1'$ peak is greater than that of $\omega_2'$ 
in the spectrum of bubble 1 and vice versa and the amplitude of the
peak corresponding to the frequency of a neighbouring bubble decreases
with the distance between them.

Analysis of Eqs~\eqref{eqr12} and \eqref{eqr22} can be performed
following the procedure outlined in \cite{Mak21}. However, it suffices
here just to note that the right-hand sides of these equations contain
quadratic terms involving $r_{11}$ and $r_{12}$ and their
derivatives. Therefore, in addition to the harmonic components with
frequencies $\omega'_{1,2}$ solutions of Eqs~\eqref{eqr12} and
\eqref{eqr22} will include steady and periodic terms with frequencies
equal to all possible pair-wise sums and differences of
$\omega'_{1,2}$ and $\Omega$: $\omega'_{1,2}\pm\omega'_{2,1}$,
$\Omega\pm\omega'_{1,2}$, $2\omega_{1,2}'$ and $2\Omega$. In the
AFC context, the frequency spectrum centred at $\Omega$ is important,
that is spectral lines $\Omega-\omega'_1(1+\Delta)$,
$\Omega-\omega'_1$, $\Omega$, $\Omega+\omega'_1$, 
$\Omega+\omega'_1(1+\Delta)$, where
\[\Delta=\frac{\omega'_2-\omega'_1}{\omega'_1}\approx
  \dfrac{1}{Q_2}-1+\frac{1}{2}\frac{1+Q^2_2}{1-Q^2_2}\zeta_{12}^2
  =1+\frac{5}{6}\zeta_{12}^2\,.\]
For an ideal AFC, $\Delta=1$, which is the case in the limit
$\zeta_{12}\to0$ of distant bubbles. It also follows from the
above expression that to keep the spectral non-uniformity of a
bubble-based AFC within 5\% it is sufficient to ensure that no
bubbles in a cluster approach each other closer than about 2 diameters 
of the largest bubble. 

To assess the robustness of a bubble-based AFC we show that the
attractive secondary Bjerknes force acting between two distant bubbles
is negligible in the AFC conditions. The expression for such a force
arising is given, for example, by Eq.~(2.5) in
\cite{Doinikov_book}. Scaled with $\rho{\omega^*}^2R_{10}^4$ and with
formal parameter $\epsilon$ set to 1 it reads
\begin{equation}
  F'_{sB}=-4\pi\zeta_{12}^2\Q_2^3\omega^2\langle r_{11}\ddot
  r_{21}\rangle\,,\label{ndbjerk}
\end{equation}
where the angle brackets denote time averaging. Substituting
expressions (\ref{eq:1order1}) and (\ref{eq:2order1}) into
Eq.~\eqref{ndbjerk} and keeping only the largest terms in each group
in the limits of $\Omega\to\infty$ and $\zeta_{12}\to0$ leads to
\begin{equation}
  F'_{sB}=\dfrac{2\pi\zeta_{12}^2\Q_2^3\omega^2}
  {(\Q_2^2-1)^2+4\Q_2^3\zeta_{12}^2}(F'_{sBn}+F'_{sBu}+F'_{sBun})\,,
\end{equation}
where
\begin{eqnarray*}
  F'_{sBn}&=&\left(\Q_2^2-1\right)
              \left(b^2-\Q_2\left(a^2+c^2-\Q_2d^2\right)\right)\zeta_{12}
          +2\left(ab\left(\Q_2^2+1\right)+2cd\Q_2^2\right)\Q_2\zeta_{12}^2\,,\\
  F'_{sBu}&=&\dfrac{\M_e^2}{\omega^4\Omega^2}\left[
              \dfrac{\left(\Q_2^2-1\right)^2}{\Q_2}
             +\left(5-2\Q_2^2+5\Q_2^4\right)\zeta_{12}^2\right]\,,\\
  F'_{sBun}&=&2\dfrac{\M_e\Q_2}{\omega^2\Omega}
               \left[\left(\Q_2^2-1\right)(d-c)\zeta_{12}
               +\left(\Q_2^2+1\right)(d+c\Q_2)\zeta_{12}^2\right]
\end{eqnarray*}
are the Bjerknes force components due to natural bubble oscillations,
ultrasound forcing and the interaction between the two,
respectively. By definition, $\Q_2<1$ and in the reference experiment
\cite{Mak21}, $\M_e/\omega^2\sim2.6\times10^{-2}$, $\Omega\sim16$
and $\zeta_{12}^2\lesssim0.04$. Therefore, we conclude that the secondary 
Bjerknes force is small at the typical driving frequencies used in the
generation of bubble-based AFCs away from bubble resonances. This
provides an opportunity for measuring the acoustic bubble response and
recording the resulting signals for AFC applications before bubble
oscillations become affected by their aggregation.

\section{Acoustic frequency combs generation using vibrations of
  liquid drops}
\begin{figure}[t]
  \centering 
  \includegraphics[width=12.0 cm]{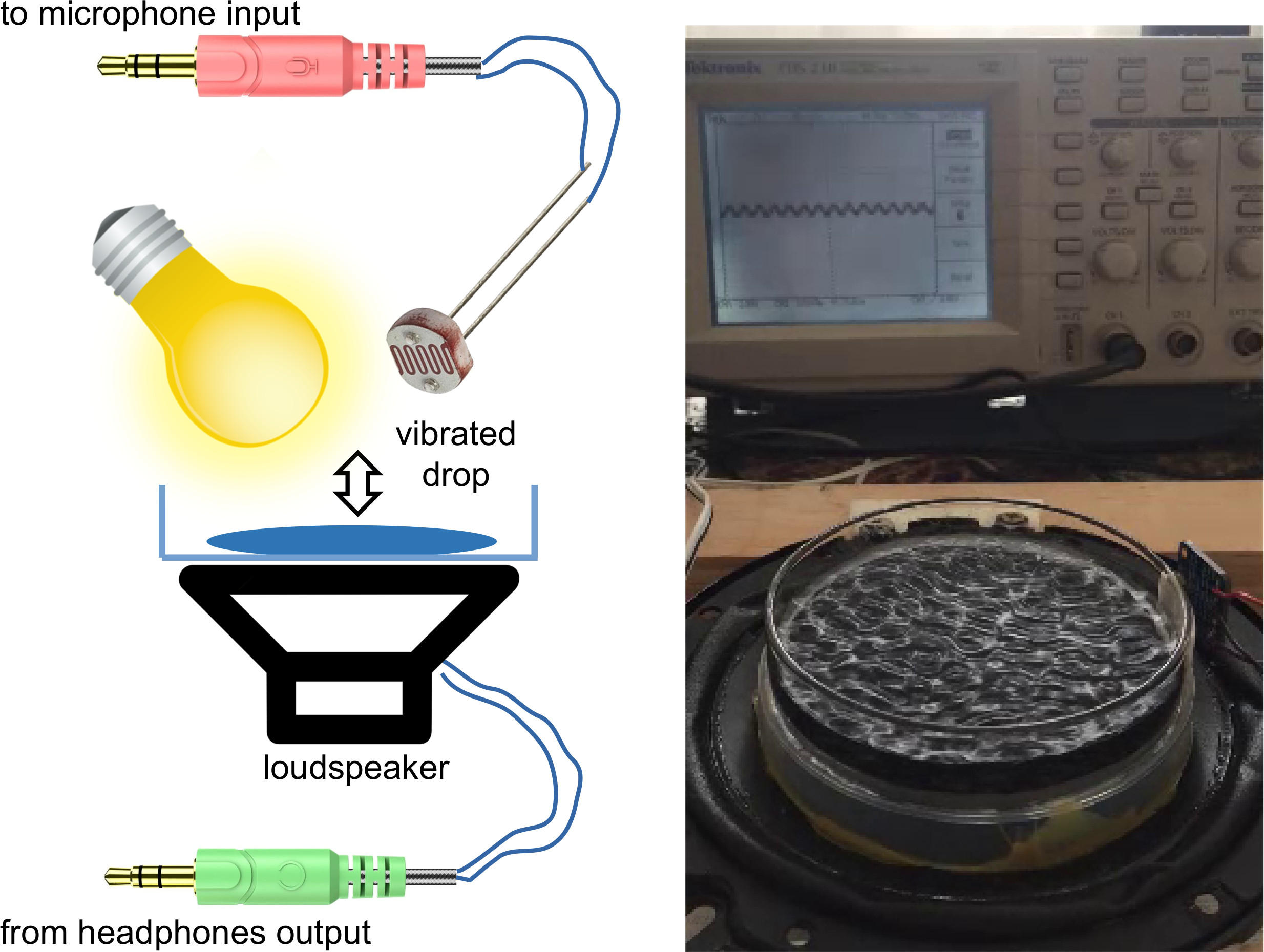}
  \caption{(Left) Schematic of a technically simple experimental setup 
    consisting of a low-frequency loudspeaker connected via a power 
    amplifier to the headphone output of a laptop computer. A Petri 
    dish is glued to the loudspeaker. An audio signal is produced by a 
    tone generator. The fluid surface is illuminated by a light source 
    and Faraday waves are detected using a photoresistor that is 
    connected either to an oscilloscope or to the microphone input of 
    the laptop computer with a pre-installed audio signal processing 
    software. 
    (Right) Photograph of Faraday waves on the surface of water 
    contained in a Petri dish glued to a vibrating loudspeaker. 
    \label{Fig16}}
\end{figure}
In this section, we discuss similarities between nonlinear acoustic
properties of liquid drops and gas bubbles in the context of AFC
generation. The natural tendency to minimisation of a surface tension
energy explains the spherical shape of undisturbed bubbles. The
surface tension also defines the shape of liquid drops that, despite
being easily deformable, tend to assume a spherical shape
also. Subsequently, similar to gas bubbles liquid drops can oscillate
in response to an acoustic forcing and develop nonlinear behaviour
analogous to that of oscillating gas bubbles \cite{Tsa83, Mak18, Mak19}.

More specifically, natural oscillations of a liquid drop are driven by
capillary forces competing with inertia of the liquid \cite{deG04}.
This leads to rich physical behaviours that warrant considering liquid
drop studies as independent and presently burgeoning field of research
\cite{Yos96, Uwe10, Puc11, Puc13, Bla13, Bos1, Bos2, Eba15, Hem15,
  Dah16, Maa16, Kam16, Chi17, Pot18, Tapio, Mak19_PRE}. An important
setup in this field is parametrically excited waves on the surface of
a vertically vibrated liquid drops, the phenomenon originally observed
on the surface of a liquid layer in 1831 by Faraday. Such waves have
become a paradigmatic example of a nonlinear wave system that exhibits
complex dynamics including periodic \cite{Ben54}, quasi-periodic 
\cite{Hen90, Mil84, Jia96} and chaotic behaviour \cite{Sha12, Pun09,
Xia12, Tur08}. Recent studies have opened new frontiers for potential
applications of Faraday waves extending beyond fluid dynamics. For
example, in photonics they have been used to generate a special kind
of OFCs \cite{Tar16}.

Physical processes behind the excitation of Faraday waves in liquids
are conceptually similar to mechanical vibrations used to generated
phononic FCs (Sec.~\ref{sec:4}). Therefore, it is plausible to assume
that Faraday waves in liquids can also be employed to generate
AFCs. However, before we discuss the relevant results we assess
several key characteristics of potential Faraday wave-based
FCs.

In liquids, the restoring force arises from surface tension
\cite{deG04}. The speed of capillary waves in a liquid is three orders
of magnitude smaller than that of acoustic pressure waves
\cite{Mak19}. As a result the frequency of capillary oscillations is
about three orders of magnitude smaller than that of an acoustic wave
mode. These observations imply that the FC generation using Faraday
waves in liquids would result in FC spectra with small inter-peak
distance of order of several tens of Hz that can find
applications, for example, in underwater acoustics \cite{Wu19}.
Moreover, capillary oscillations of liquid drops can be excited using
non-mechanical techniques such as electrowetting, where a sessile
drop oscillates when an alternating voltage is applied to it via a contact
electrode. The oscillations result from a time-varying electrical
force acting on the three-phase contact line. They lead to 
resonances that occur at certain frequencies of the applied
alternating electric signal \cite{Oh08}. While for mm-sized liquid drops such
resonance frequencies can be in the range from 30 to 300\,Hz, using
room-temperature liquid metal alloy nanodrops one can increase them
to several GHz \cite{Mak17_PRA}, which is beneficial for AFC
applications. We will return to the discussion of the relevant
physical properties of liquid metal drops in Sec.~\ref{sec:liquid_metals}. 

However, relatively low viscosity of common liquids such as water
implies that Faraday surface waves can be excited using low-amplitude 
vertical vibrations produced by inexpensive and readily available components
such as loudspeakers and piezoelectric transducers (Fig.~\ref{Fig16}) that 
are much simpler than any equipment used in electrowetting experiments
\cite{Oh08}. As discussed in \cite{Mak20_worms, Pot20, Pot21}, enhancements
of the basic setup used to investigate Faraday waves in oscillating drops
also include common elements such as diode lasers, measurement-grade
photodetector and accelerometer, which nevertheless revealed a number
of intriguing physical processes that can also help understand the 
behaviour of smaller drops oscillating at much higher frequencies
\cite{Tsa17}.

\subsection{Experimental demonstration of AFC generation using Faraday
  waves} 
To demonstrate the plausibility of AFC generation using Faraday waves 
in liquids, in the experiment of \cite{Mak19_SPIE} a red laser diode
(650\,nm wavelength, 0.5\,mW power) with a highly-divergent beam profile
was used as the source of light. A loudspeaker (3\,W, 20\,Hz-15\,kHz) was
driven via an audio amplifier by a pure sinusoidal signal $f=100$\,Hz.
One end of a cardboard cylinder (height 20\,cm, radius 4.2\,cm) was fixed
above the loudspeaker and the other end was covered with a 0.5\,mm thick
black Teflon membrane. A pancake-like drop of pure alcohol (95\% v/v
ethanol) was placed on top of the membrane. The thickness of the liquid layer
was 2\,mm. A photodetector with a frequency response covering the entire
frequency range of a loudspeaker was used to receive light reflected from
the liquid surface.

\begin{figure}[t]
  \centering
  \includegraphics[width=12.0 cm]{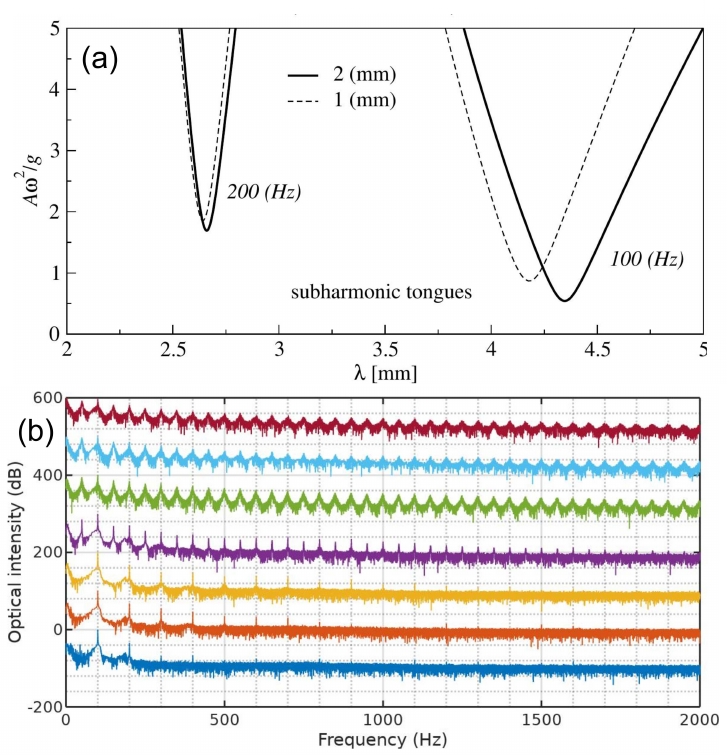}
  \caption{\textbf{(a)}~Stability diagram for the $h=2$\,mm and
    $h=1$\,mm deep ethanol layers showing two subharmonic Faraday 
    tongues. The oscillation wave frequency is $f_F=50$\,Hz, the
    half of the vertical vibration frequency $f=100$\,Hz. 
    \textbf{(b)}~Experimental optical spectra for seven gradually
    increasing (from bottom to top) amplitudes of the vibration signal
    with $f=100$\,Hz. The depth of the ethanol layer is $h=2$\,mm. All
    spectra are vertically offset by 100\,dB. The three lowest spectra
    are dominated by the vibration frequency $f=100$\,Hz and its
    higher-order harmonics $nf$ ($n=2,3,\dots$) that appear due to
    nonlinear acoustic effects in ethanol. The fourth and so on
    spectra are dominated by the Faraday wave at $f_F=50$\,Hz and its
    higher-order harmonics. Note that all peaks have a characteristic
    triangular shape (see the main text). The equal spacing between
    them allows using the spectra as AFCs. Reproduced from \cite{Mak19_SPIE}.  
    with permission of SPIE and the authors of the publication.\label{Fig17}}
\end{figure}

The classical nonlinear standing Faraday waves appear on the surface
of a horizontally extended fluid in a vertically vibrating
container. When the normalised vibration amplitude $A\omega^2/g$ 
($\omega=2\pi f$, $f$ is the vibration frequency and $g$ is the
gravity acceleration) exceeds the critical value, a flat fluid surface becomes
unstable and subharmonic surface waves oscillating at the frequency
$f/2$ are formed. Figure~\ref{Fig17}(a) shows the stability diagram
for $h=2$\,mm and $h=1$\,mm deep ethanol layers, where the
so-called subharmonic Faraday tongues that correspond to the neutral
stability of perturbation with wavelength $\lambda$ and half of the
driving frequency $f/2$ can be seen \cite{Rab89}. The flat fluid
surface is linearly unstable above the boundary of the tongues. For
$f=100$\,Hz (200\,Hz) and $h=2$\,mm, Faraday wave wavelength 
is $\lambda=4.34$\,mm (2.66\,mm) at $A\omega^2/g=0.53$~(1.7). For
$f=100$\,Hz and $h=2$\,mm, a high-speed camera was used to estimate
wavelength $\lambda\approx4.5$\,mm, which is in good agreement with
the theoretical prediction.

However, in the experiment of \cite{Mak19_SPIE} the fluid forms a
pancake-shaped liquid drop on the surface of a solid membrane. The
formation of Faraday waves in this case is qualitatively
different. When such a drop is vertically vibrated with a small
amplitude, the capillary surface waves are excited at the edge of a
drop, i.e.~at the contact line between the fluid and the
membrane. Similarly to a classical damped harmonic oscillator, the
frequency of the excited waves is identical to the vibration 
frequency $f$. This result is confirmed in Fig.~\ref{Fig17}(b) (see
the three lowest spectra). As the amplitude increases remaining below
the critical value, new peaks appear at the harmonic frequencies $nf$
($n=2, 3, \dots$). In agreement with the theory, these peaks are due
to nonlinear-acoustic effects in ethanol triggered by the incident
acoustic wave $f=100$\,Hz. The onset of Faraday waves in
finite-volume liquid drops is associated with the period-doubling
bifurcation \cite{Pot18}. When the amplitude of the 100\,Hz wave
reaches the Faraday instability threshold, surface Faraday waves are
excited at $f_F=f/2=50$\,Hz. The nonlinearity of these waves is so
strong that one can observe many higher-order harmonics
$nf_F$. The height of the peaks associated with Faraday waves
is much larger than that of at driving $100$\,Hz frequency of the
acoustic wave and it has a characteristic triangular shape. This shape
is a signature of extreme nonlinearities observed in fluid-mechanical
systems \cite{Pun09} leading to the formation of capillary rogue waves
\cite{Sha10}, oscillons\cite{Sha12} and solitons \cite{Raj11}.

\begin{figure}[t]
  \centering
  \includegraphics[width=12.0 cm]{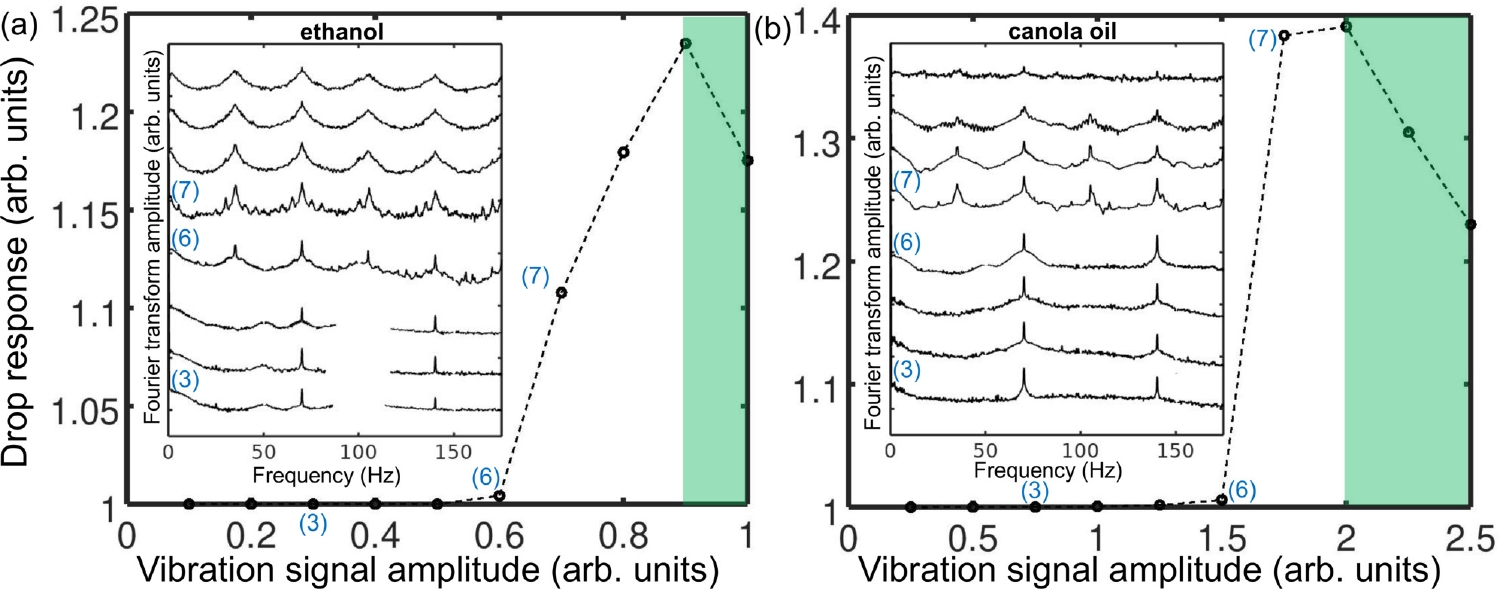}
  \caption{\textbf{(a)}~Experimental average response of an ethanol
    drop subjected to vertical vibration at 70\,Hz plotted as a
    function of the vibration amplitude. The inset shows the power
    spectra obtained by Fourier-transforming the measured signals. The
    parenthetical labels relate the spectra to the experimental points
    in the main panel.
    \textbf{(b)}~Experimental average response of a canola oil drop 
    subjected to vertical vibration at 70\,Hz plotted as a function 
    of the vibration amplitude. The modulation sidebands (the
    spectrum label 7) are present in the ethanol drop spectra and
    absent in the spectra of a canola oil drop. The shaded regions
    in the main panels correspond to the chaotic oscillations
    resulting in strong diffuse scattering leading to a decrease
    in the optical intensity of the detected signal. Reproduced
    from \cite{Mak19_PRE}. Copyright 2019 by the American Physical
    Society.\label{Fig18}}
\end{figure}

To understand the physics behind the changing peak shape, a more
detailed experiment was conducted in \cite{Mak19_PRE} using smaller
pancake-like drops of ethanol and canola oil (Fig.~\ref{Fig18}). The
observations revealed a number of peculiarities relevant to the
generation of AFC. Their theoretical explanation is given in the
following section.

\subsection{Existence conditions for Faraday-wave-based acoustic
  frequency combs} 
Here, we summarise the results of the analysis reported in
\cite{Mak19_PRE} with a special focus on the physical conditions that
are favourable for the generation of Faraday wave-based AFCs.   
To start with, we note that modelling the dynamics of a liquid drop
that rests on a solid plate requires a correct description of the
contact line motion. To avoid the well-known hydrodynamic
singularities at the true contact line \cite{Huh71}, a standard
regularisation method is used that is based on the assumption that a
molecularly thin precursor film exists beneath the drop and covers the
entire area of the solid plate \cite{deG04}. In this case, the
equilibrium contact angle can be determined by balancing the pressure
in the precursor film and in the drop. For small contact angles, the
total pressure can be written in terms of the drop thickness $h(x,y)$
\cite{deG04} 
\begin{eqnarray}
  P=-\sigma(\partial_{xx}+\partial_{yy})h+\rho gh-\Pi(h)\,,\label{eq1}
\end{eqnarray}
where $\sigma$ and $\rho$ are the surface tension and the fluid
density and $\Pi(h)$ is the disjoining pressure that describes
the long-range van der Waals forces giving rise to the formation
of the precursor film. In general, the function $\Pi(h)$ that
allows the existence of a steady drop with a non-zero contact
angle is given by \cite{Sch98}
\begin{eqnarray}
  \Pi(h)=\frac{B}{h_{\infty}^n}\left(\frac{h_{\infty}^n}{h^n}
  -\frac{h_{\infty}^m}{h^m}\right)\,,\label{dp}
\end{eqnarray}
where $B$ is a constant, $h_{\infty}$ is the thickness of the 
precursor film and $m, n$ are some integer numbers. Following
\cite{Uwe10, Pot18} and choosing $n=6$ and $m=3$,
the disjoining pressure Eq.~\eqref{dp} can be written in
terms of the Hamaker constant $A_H$, a constant that accounts
for van der Waals interaction between two materials, as
\begin{eqnarray}
  \Pi(h)=\frac{A_H}{h^3}\left(\frac{h_{\infty}^3}{h^3}-1\right)\,.\label{dp1}
\end{eqnarray}
The equilibrium contact angle $\theta\ll1$ is related to $A_H$
and $h_\infty$ via \cite{Sch98}
\begin{eqnarray}
  A_H\approx\frac{5h_\infty^2}{3}\sigma\theta^2\,.\label{dp2}
\end{eqnarray}
Subsequently, we adopt a simplified version of the model proposed in 
\cite{Pot18}, where we only take into account long-wave deformations
of the liquid-gas interface assuming a quadratic dependence of the
horizontal fluid velocity ${\bm u}(x,y,z)$ on the vertical coordinate
$z$ 
\begin{eqnarray}
  {\bm u}={\bm\Phi}\left(\frac{z^2}{2}-hz\right)\,.\label{eq2}
\end{eqnarray}
This expression satisfies the boundary conditions at the solid plate
${\bm u}(z=0)=0$ and at the liquid-gas interface $\partial_z{\bm
  u}(z=h)=0$ for some arbitrary function ${\bm\Phi}(x,y)$. The flow
field across the layer ${\bm q}=\int_0^h {\bm u}\,dz$ can be expressed
in terms of ${\bm \Phi}$
\begin{eqnarray}
  {\bm q}=-{\bm \Phi}\frac{h^3}{3}\,.\label{eq3}
\end{eqnarray}
Navier-Stokes equation in the long-wave approximation is
\begin{eqnarray}
  \rho\left(\partial_t {\bm u}+{\bm\nabla}({\bm u}^2)
  +\partial_z({\bm u}w)\right)=\mu\partial_{zz}{\bm u}
  -{\bm\nabla}P\,,\label{eq4}
\end{eqnarray}
where $\mu$ is the dynamic viscosity, $w$ is the vertical component of
the velocity and $P$ is the pressure. Integrating Eq.~\eqref{eq4} over
$z$ and using the kinematic boundary condition
$\partial_t h +({\bm u}\cdot {\bm \nabla} h) = w$, we obtain 
\begin{eqnarray}
  \rho\left[\partial_t {\bm q}+\frac{6}{5}{\bm \nabla}
  \cdot\left(\frac{{\bm q}\otimes{\bm q}}{h}\right)\right]
  &=&-\frac{3\mu {\bm q}}{h^2}-h{\bm \nabla} P,\nonumber\\
  \partial_t h + ({\bm \nabla}\cdot {\bm q}) &=&0\,,\label{eq5}
\end{eqnarray}
where ${\bm q}\otimes{\bm q}$ denotes the matrix product.

When a drop is supported by a solid plate that vibrates vertically
with amplitude $A_0$ and frequency $\Omega$, Eqs~\eqref{eq5}
are valid in the frame co-moving with the plate. The pressure $P$
is taken from Eq.~\eqref{eq1} with $g$ replaced by
$g(1+a\cos{\Omega t})$, where $a=A_0\Omega^2/g$ is the dimensionless
vibration amplitude. The validity of system Eqs~\eqref{eq5} is
restricted to small contact angles and small variations of
the drop height $h$. In addition, the characteristic horizontal
deformation wavelength $\lambda$ must be larger than the length
of the viscous boundary layer $l=\sqrt{2\mu/(\rho\Omega)}$
associated with the vibration frequency $\Omega$ \cite{Bes16}.
In what follows, we consider the simplest possible case of a one-dimensional
liquid drop, whose dynamics is described by Eqs.\,(\ref{eq5}) with $h(x,t)$
and one-dimensional fluid flux $q(x,t)$.

\subsection{Linear response and higher harmonics}
In the absence of external driving, i.e. when $a=0$, Eqs.(\ref{eq5}) admits
a steady-state solution $h_s(x)$, characterised by zero fluid flux $q(x,t)=0$.
In an unbounded domain, this solution satisfies
$\lim_{x\rightarrow \pm \infty}h_s(x)=h_\infty$ and corresponds to an equilibrium
liquid drop resting on the precursor film of thickness $h_\infty$. Drops with a
sufficiently large volume (excluding the volume of the precursor film) are
flattened by the gravity, resembling a cross-section of a pancake. The height
$h_0$ of the drop is much larger than the precursor film thickness $h_\infty$
and the volume of the drop determines its width $W$. 

In response to a weak vibration $a\ll 1$, the drop develops the so-called harmonic
small-amplitude capillary waves on its surface, whose oscillation frequency is
identical to the vibration frequency $\Omega$. The exact form of the temporal
response of the drop to a weak external vibration can be determined by
linearising Eqs.\,(\ref{eq5}) about a flat upper cap of the drop $h=h_0+\tilde{h}(x,t)$,
where $\tilde{h}\ll h_0$ is a small deformation amplitude. As it was shown in
\cite{Mak19_PRE}, the resulting linearised equation is given by the damped-driven
Mathieu equation
\begin{eqnarray}
  \label{eq_mathieu}
 \partial_{tt} \tilde{h} +\frac{3\mu}{\rho h_0^2}\partial_t \tilde{h} + h_0\left(\frac{\sigma}{\rho} \partial_x^4+g(1+a\cos{\Omega t})\partial_x^2\right) \tilde{h}=p_e(x,t)\,,
  \end{eqnarray}
where $p_e(x,t)$ denotes the excess pressure, generated by the left and the right oscillating drop edges.

Next we observe that $p_e(x,t)$ oscillates with the driving frequency $\Omega$
and expand $\tilde{h}=a\tilde{h}^{(0)}+a^2\tilde{h}^{(1)}+a^3\tilde{h}^{(2)}+\dots$.
The analysis conducted in \cite{Mak19_PRE} based on Eq.\,(\ref{eq_mathieu}) reveals
that the leading order response $\tilde{h}^{(0)}$ is harmonic, i.e.~$\tilde{h}^{(0)}$
oscillates with the driving frequency $\Omega$. However, the higher-order terms
$\tilde{h}^{(1,2,\dots)}$ contain higher-order harmonics of the driving signal.
It is therefore evident that already in the linear regime, the temporal spectrum
of the drop response can be used as a frequency comb with delta-like peaks at
frequencies $n\Omega$, $(n=1,2,3,\dots)$.

\subsection{Nonlinear response and the amplitude modulation}
%%%
To investigate the nonlinear response of the drop and to further identify 
a regime that would be favourable for the AFC generation, we employ the
continuation method, where Eqs~\eqref{eq5} are 
solved numerically for gradually varying (increasing or decreasing)
values of the vibration amplitude $a$: $a_n=n\Delta$, $n=0,1,2,3,\dots$
with a fixed step $\Delta=0.0015$. For each value of $n$
Eqs~\eqref{eq5} are integrated over the time interval that is 
equivalent to $1000$ oscillation cycles. The solution from the
$(n-1)$th  run is taken as the initial condition for the $n$th run. To
artificially recreate the experimental conditions reported in 
\cite{Mak19_PRE}, a small-amplitude white noise $\xi(x,t)$ is added to
the second of Eqs~\eqref{eq5}. 

\begin{figure}[t]
  \centering  
  \includegraphics[width=12.0 cm]{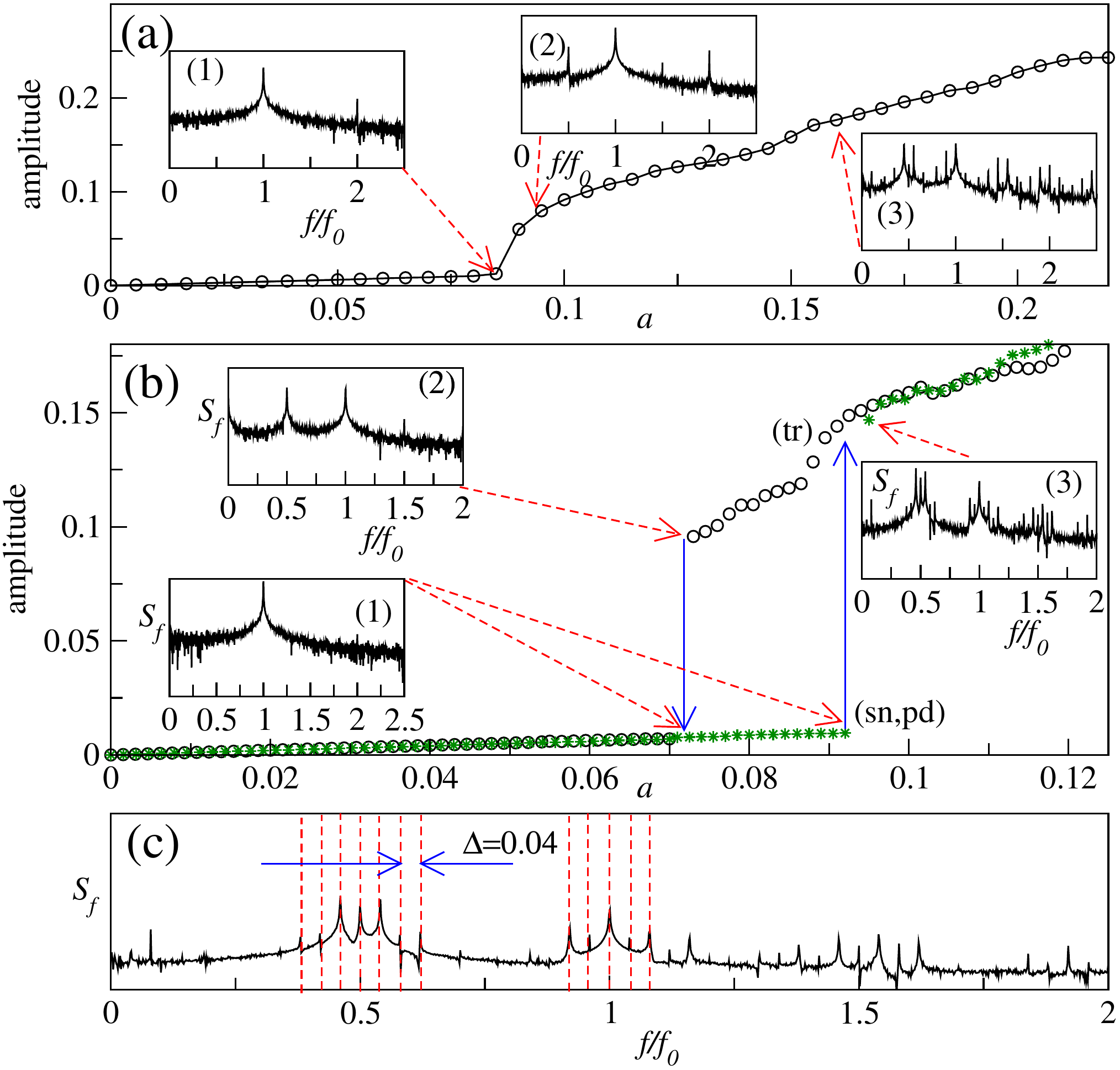}
  \caption{Average response amplitude of an ethanol drop vibrated at   
    the frequencies \textbf{(a)}~$f$\,=\,21 and \textbf{(b)}~28\,Hz
    as a function of the gradually increasing amplitude $a$. Harmonic
    waves lose their stability \textbf{(a)}~via a supercritical
    period-doubling bifurcation (pd) at $a=0.08$ and \textbf{(b)}
    via a torus bifurcation at $a=0.09$. The modulational instability
    sets in via a torus bifurcation (tr) on the subharmonic branch. At  
    $f=28$\,Hz one observes the formation of a hysteresis loop  
    consisting of a gradually increasing (asterisk) and gradually  
    decreasing (circle) branches, where sn and tr correspond to  
    the saddle-node and torus bifurcations, respectively. 
    \textbf{(c)}~Close-up of spectrum $S_f$ (3) in panel (b) in
    the regime of the developed modulational instability. The dashed
    vertical lines highlight the location of equally distant delta
    peaks with the scaled inter-peak distance of $f/f_0=\Delta=0.04$.
        \label{Fig19}}
\end{figure}

The deformation $\delta h(t)$ of the drop surface at $x=0$ and its power
spectrum $S_f$ are used to characterise the temporal response of a fixed-volume
ethanol drop vibrated with different frequencies $f$. In Fig.~\ref{Fig19} the
amplitude of $\delta h(t) $ in the units of the drop height $h_0$ is
shown as a function of $a$ for a fixed-volume ethanol drop vibrated at 21\,Hz (Fig.~\ref{Fig19}(a))
and 28\,Hz (Fig.~\ref{Fig19}(b)). 

The characteristics of the drop response are highly sensitive
to the relationship between the Faraday wavelength $\lambda_F$ and the
horizontal size of the drop $W$. Since in the relevant experiment the
volume of the drop was fixed, in the model the vibration frequency $f$
was varied resulting in a change of the dimensional wavelength of
Faraday waves. Thus, at the vibration frequency 21\,Hz, we observe
the onset of the sub-harmonic Faraday waves via a super-critical
period-doubling bifurcation that occurs at amplitude $a=0.08$, as
shown in Fig.~\ref{Fig19}(a). The newly established mixed state has a high
degree of temporal order characterised by the sharp subharmonic and
harmonic peaks in the frequency spectrum. At around $a=0.15$ the mixed
state undergoes a secondary bifurcation that corresponds to the
modulation instability of wave amplitude. While from the dynamical
theory point of view this instability corresponds to a torus
bifurcation (tr), the resulting spectrum develops a number of
equidistant frequency peaks that can be used as an AFC. 

A qualitatively different scenario was obtained for $f=28$\,Hz, see
Fig.~\ref{Fig19}(b), where the primary instability of harmonic waves is
most likely a torus bifurcation. When $a$ is increased with step
$\Delta$, harmonic waves follow the branch shown by asterisks. They
lose stability at around $a=0.092$ (labels (sn) and (tr)). The response
amplitude jumps sharply and no subharmonic peak is visible in the
temporal spectrum thereby speaking in favour of a torus or saddle-node
bifurcation. A new state corresponds to the modulated Faraday waves,
which is confirmed by the presence of frequency sidebands in the
temporal spectrum. These are required for the AFC
generation. Moreover, as the vibration amplitude $a$ is decreased from
$a=0.12$, the response amplitude follows the branch shown by circles
that stretches to $a=0.07$. At $a=0.087$ (label~(tr)), the modulated 
Faraday waves are replaced with non-modulated standing waves via a
reversed torus bifurcation. Yet, for $a<0.07$ (label~(sn)) the branch
of standing non-modulated Faraday waves do not exist and the drop
response is harmonic. The subcriticality of bifurcations results in
a hysteresis loop shown by the thick blue arrows in Fig.~\ref{Fig19}(b).

The physical mechanism behind the formation of the AFC in the regime of
the modulational instability is the nonlinear mixing of the three
frequencies: the driving frequency $f$, the sub-harmonic frequency of
the Faraday waves $f/2$ and the much smaller frequency $f_m \ll f$
which corresponds to the amplitude modulation. A close-up of the temporal
spectrum (3) from Fig.~\ref{Fig19}(b) in the regime of the developed
modulational instability is shown in Fig.~\ref{Fig19}(c). The dashed
vertical lines highlight the location of the delta peaks in the power
spectrum of the response of the drop. The peaks appear as equidistant,
with a scaled inter-peak interval of $f/f_0=\Delta =0.04$, where
$f_0=28$ Hz is the driving frequency.

\begin{figure}[t]
  \centering 
  \includegraphics[width=12.0 cm]{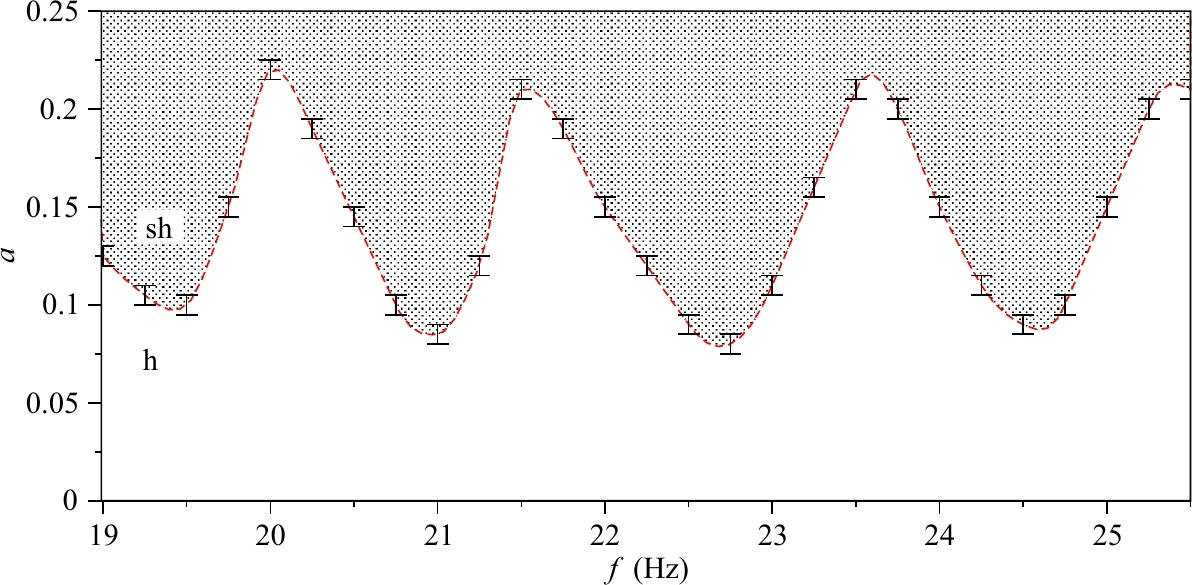}
  \caption{Frequency-amplitude phase diagram of the harmonic (h) and 
    subharmonic (sh) responses for a fixed volume ethanol drop 
    vertically vibrated with the frequency $f$. The response is 
    subharmonic in the shaded area, where the generation of AFCs has 
    been observed. Reproduced from \cite{Mak19_PRE}. Copyright 2019
    by the American Physical Society.\label{Fig20}}
\end{figure}

As follows from the discussion Fig.~\ref{Fig19} the AFC generation is
possible only when the drop size and the frequency of its vertical
vibrations fall within specific ranges. This observation was confirmed
experimentally as shown in Fig.~\ref{Fig18} although slightly higher
vibration frequencies were used. To visualise the existence range of 
AFCs, in Fig.~\ref{Fig20} we plot the boundary between harmonic and
subharmonic responses in the frequency-amplitude plane for a fixed
volume ethanol drop. The boundary is obtained by gradually increasing
the vibration amplitude $a$ at a fixed frequency $f$. The response is 
subharmonic in the shaded area, where the AFC generation has been
observed. A high sensitivity of the Faraday instability threshold to
frequency variation is explained by a simple geometric commensurability
condition: if the horizontal drop size is an integer multiple of a
half of the Faraday wavelength $\lambda_F/2=\pi$, then subharmonic
Faraday waves are easier to excite on the drop surface. 

Finally, note that both theory and experiment in \cite{Mak19_PRE}
demonstrate that in highly viscous fluids the modulation instability
of subharmonic Faraday waves is pushed towards larger values of
vibration amplitude, thereby affecting the AFC existence conditions
outlined above since in this case the primary instability of harmonic
waves is dominated by a period-doubling bifurcation and modulation
sidebands are not found in the frequency spectrum. Hence, we conclude
that highly viscous liquids would be less suitable for the generation of
Faraday wave-based AFCs.

\subsection{Faraday waves liquid-metal drops at room temperature\label{sec:liquid_metals}}
In this section, we discuss the perspectives of the AFC generation using
nonlinear acoustic properties of novel non-toxic, room-temperature
liquid-metal materials notable for their unusual electronic, fluid mechanical
and optical properties \cite{Dic17, Dae18, Rei19, Neu20, Boy20}.
For example, in \cite{Rei19} eutectic gallium-indium alloy (EGaIn, 75\% Ga
25\% In by weight, melting point approximately $-15.5^\circ$C)
spherical nanoparticles suspended in ethanol were investigated and 
their strong plasmonic resonances in the UV spectral range were 
demonstrated. Previously, nanoparticle-like liquid droplets were 
investigated as ultra-small mechanical resonators that can be used
to generate FC-like signals \cite{Maa16, Dah16}. However, liquids
used in those studies did not support any plasmon mode because they
did not possess the conductivity of metals. On the contrary,
liquid metals combine the properties of both liquids and
metals. Therefore, their drops can oscillate like pure liquid and
exhibit plasmonic properties at the same time. This was demonstrated
in \cite{Mak17_PRA}. Liquid metals also change their shape easily. For
example, it was suggested in \cite{Boy20} that an acoustically driven 
oscillating gas bubble located above a liquid-metal layer can modify the
liquid-metal surface creating a parabolic-like micro-mirror that can
focus UV light into a beam.

Here, we discuss another opportunity that liquid metals offer 
in the field of AFC generation using surface Faraday waves. EGaIn and 
similar Ga-based alloys have the following relevant material parameters: 
surface tension $\sigma=624$\,mN/m, dynamic viscosity
$\mu=1.99\times$10$^{-3}$\,Pa\,s and density $\rho=6280$\,kg/m$^3$,
which we use in calculations supporting this discussion. However, the
surface of EGaIn is covered by a nanometer-thin oxide layer
\cite{Dic17, Rei19} that does not dissolve into the bulk metal and is
also technologically important for the formation of stable
drops. Furthermore, the surface tension of EGaIn is reversibly
changeable from 0.624\,N/m to 0.07\,N/m (approximately the 
surface tension of water) when an external voltage is applied to a liquid
metal layers \cite{Dic17, Dae18}. The use of these properties creates
new opportunities for studies of Faraday waves and relevant nonlinear
phenomena in liquid metals.

For example, the onset of Faraday waves on the surface of a large liquid 
pancake-like drop made of pure Ga was observed in \cite{unpubl,
  conf_Madrid}. In that experiment, the frequency of vertical
vibration was varied from 40 to 300\,Hz. It was found that conditions
for the generation of FC-like signals could be satisfied at much lower
vibration amplitudes than those in the experiments involving ethanol and
canola oil drops of similar dimensions \cite{Mak19_PRE}. These
observations are in good agreement with the result of the analysis that we
discuss below.
\begin{figure}[t]
  \centering
  \includegraphics[width=12.0 cm]{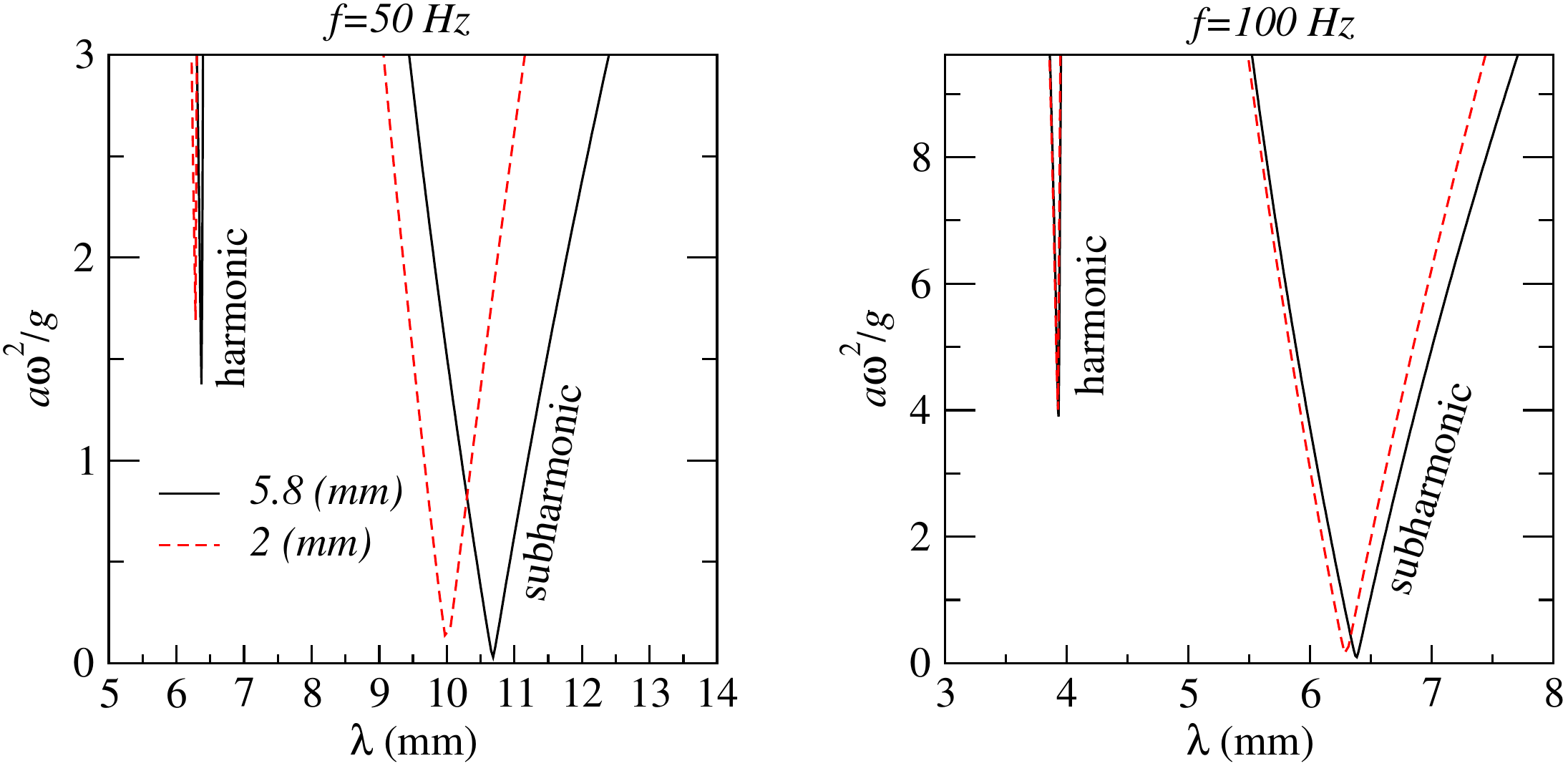}
  \caption{Stability diagram for $h=5.6$\,mm and $h = 2$\,mm deep
    liquid metal layers showing two subharmonic Faraday tongues. The
    oscillation frequency of Faraday waves is half of the vertical
    vibration frequency $f=50$\,Hz (left) and 100\,Hz (right).
    \label{Fig21}}
\end{figure}    

Note that the thickness $h$ of the pancake-like drop of the liquid
metal created on the surface of a vertically vibrated plate is not a
free parameter but is related to the contact angle $\alpha$ as
\begin{equation}
  (1/2)\rho gh^2=\sigma[1-cos(\alpha)]\,.\label{eq:eq_liq1}
\end{equation}
For example, using a typical value $\alpha=160^\circ$ \cite{Liu12} one
obtains $h=5.8$\,mm. Figure~\ref{Fig21} shows the stability diagram
for a liquid metal that demonstrates that the Faraday threshold for
this material is much lower than that of ethanol in
Fig.~\ref{Fig17}(a). Furthermore, the diagram predicts that a higher 
amplitudes $a$ the excitation of harmonic perturbations is observed,
which further increases the magnitude of multiple spectrum peaks thus
making them usable for AFC applications. An intuitive physical reason
for a low Faraday threshold for a liquid metal is that the threshold
is controlled by the kinematic viscosity $\eta=\mu/\rho$, which is
estimated as $1.5\times10^{-6}$ for ethanol and $3.7\times10^{-7}$ 
for liquid metal.

Of course, the use of Faraday waves on the surface of a liquid
metal drop for AFC purposes has physical limitations. For example, it
is technically challenging to excite sub-micrometer patterns that, 
for example, would be required to enable a strong interaction
of light with the surface Faraday waves---such a wave pattern would be 
perceived by the light as an oscillating diffraction grating, which in
turn could lead to intriguing optical effects in the frequency range,
where liquid metal supports plasmon resonances \cite{Rei19}. In fact, the Faraday
instability originates from a parametric excitation of gravity-capillary waves
on the surface of a fluid. When the effect of viscosity is neglected, which is a
valid approximation for liquid metals, the dispersion relation for
gravity-capillary waves in shallow layers can be written as
\begin{equation}
  \omega^2=(gk+\sigma k^3/\rho)tanh(kh)\,,\label{eq:eq_liq2}
\end{equation}
where $\omega$ is the driving angular frequency, $h$ is the thickness of the layer
and $k=2\pi/\lambda$ is Faraday wavenumber. When $\omega\gg1$, the
wavenumber is also large so that the gravity in the dispersion
relation can be neglected.  This leads to the wavelength
\begin{equation}
  \lambda=2\pi(\sigma/(\omega^2\rho))^{1/3}\,.\label{eq:eq_liq3}
\end{equation}
If $\omega=1$\,MHz then $\lambda=14$\,$\mu$m. To excite a
sub-micrometer wavelength, which would give rise to optical
  diffraction grating-like oscillations on the liquid metal surface,
the frequency should be of order of several GHz. This presents a
significant technical challenge because it is difficult to produce a
mechanical vibration with the frequency in the GHz range. 
  However, as discussed above, in liquid metal drops Faraday waves
  could also be excited using an electrowetting technique \cite{Oh08,
    Tsa17}. As theoretically shown in \cite{Mak17_PRA}, using
  electrowetting one can excite oscillations of the liquid metal
  surface with GHz-range frequencies. Although the application of such
  results in the context of AFC generation was not the focus of the
  work \cite{Mak17_PRA}, a dramatic change in the frequency of the
  plasmon resonance due to capillary oscillations found there
  unambiguously speaks in favour of plausibility of optical wave
  modulation and appearance of sideband peaks in the optical 
  spectrum of the incident light, thus resulting in a signal that 
  could be used as an AFC.         

Thus, the application of non-toxic, room-temperature liquid-metal
alloys holds the promise to become a useful method of AFC
generation. While in many applications the use of solid-state AFCs
could be preferred, in some situations liquid-state technologies  
can be advantageous. For example, liquid metals discussed 
in this section have the potential to be used inside a living 
body to realise such important functions as sensing and drug delivery
\cite{Lu15, Kal19, Rei19}. Therefore, it is plausible that a combination of
these new approaches with an FC-based medical technology \cite{Hen18}
could become a subject of future research. 

\section{Conclusions and outlook}
There has been a significant progress in understanding the fundamental
physical processes that underpin the operation of AFCs---non-optical
counterparts of OFCs. As with OFCs that are essential in applications
that require the highest accuracy and resolution when measuring the
frequency, time, distance or molecular composition of a material using
light, AFCs have enabled similar functionalities using sound and
vibrations in practical situations where light cannot be used. Similar
to OFCs that have evolved into an independent technology soon after
their introduction, the recently proposed AFCs have already enabled
several novel spectroscopy and microscopy techniques, sensors and
medical imaging modalities. Hence, it is plausible that the field
of AFCs will rapidly grow and that this emergent technology will
find further applications in science and engineering. 

However, extra research and development work is required for the AFC
technology to fulfil its full potential. Here, we outlined some of the
important milestones that will need to be reached to achieve this goal.

Firstly, whereas AFCs are expected to operate similarly to OFCs, there
are several differences between these two techniques stemming from the
disparate frequencies of acoustic waves (as well as of vibrations and
spin waves) and light and also with physical mechanisms by which these
waves interact with media. This aspect presents numerous technological
challenges that shape research efforts in the field of AFCs. 

Secondly, the type of AFC suitable for a particular practical
application, and thus the physical mechanism of its generation,
depends on specific experimental conditions and technical
requirements. While this observation also applies to OFCs since they
can be generated using a number of different techniques, in the case
of AFCs one has a much wider choice of approaches to the comb
generation. For example, whereas generating AFC spectra with peaks of
the same magnitude would be advantageous for certain applications, 
having peaks of different heights---which is the case with several
AFCs discussed in this article---can be, in general, inconsequential
as long as the peaks are detectable and their frequencies are stable.

Furthermore, some AFCs have a smaller number of spectral peaks
than OFCs, which is feature that has been identified as being
important for a number of practical applications including phonon
lasers \cite{Gru10, Bea10} and computing \cite{Sta12}. Such AFCs
should be compared with opto-electronic OFCs that are known to also
have a small number of peaks and require special techniques for
broadening their spectral ranges \cite{Zha19_1}. The same applies,
although to a lesser degree, to Kerr OFCs generated using a cascade of
optical FWM processes \cite{Che16}. Indeed, a Kerr OFC may have just 
ten or so peaks due to intrinsically low strength of nonlinear optical
effects \cite{Mak13, Mak19}, which is in stark contrast to
conventional OFCs based on mode-locked lasers and containing hundreds
of frequency peaks \cite{Pic19}.
 
Thirdly, although a considerable attention has been paid to the
long-term temporal stability of certain kinds of AFCs and conclusions
have been drawn regarding their applicability to resolve specific
problems such as precision underwater measurement, the stability of
AFCs of the other kinds has not been investigated in great
detail. While filling this knowledge gap is likely to be one of the
future directions in the AFC research, the lack of long-term stability
may be inconsequential in some applications such as, for example,
sensing and imaging in {\it in vivo} biological environments
\cite{Mak16, Mak20_worms} that evolve in time.

Finally, we note that some of the AFCs technologies discussed in this
review article have already been patented (see, e.g.,~\cite{Roo18,
  Ans19}) and therefore they are expected to enter service in
commercial devices and systems in the nearest future.
     
%%%%%%%%%%%%%%%%%%%%%%%%%%%%%%%%%%%%%%%%%%
\vspace{6pt} 
%%%%%%%%%%%%%%%%%%%%%%%%%%%%%%%%%%%%%%%%%%

\funding{ISM has been supported by the Australian Research Council
through the Future Fellowship (FT180100343) program.} 

\acknowledgments{ISM thanks Professor Mikhail Kostylev (The University
of Western Australia) and Professor Vladislav Yakovlev (Texas A\&M
University) for invaluable discussions of the magnonic and phononic
BLS techniques, and Professor Andrew Greentree (RMIT University) for
enlightening discussions of acousto-optical effects.}

\conflictsofinterest{The authors declare no conflict of interest.} 

%%%%%%%%%%%%%%%%%%%%%%%%%%%%%%%%%%%%%%%%%%
%% Only for journal Encyclopedia
%\entrylink{The Link to this entry published on the encyclopedia platform.}

%%%%%%%%%%%%%%%%%%%%%%%%%%%%%%%%%%%%%%%%%%
\end{paracol}
%%%%%%%%%%%%%%%%%%%%%%%%%%%%%%%%%%%%%%%%%%
% To add notes in main text, please use \endnote{} and un-comment the
% codes below. 
%\begin{adjustwidth}{-5.0cm}{0cm}
%\printendnotes[custom]
%\end{adjustwidth}
%%%%%%%%%%%%%%%%%%%%%%%%%%%%%%%%%%%%%%%%%%
\reftitle{References}

% Please provide either the correct journal abbreviation
% (e.g. according to the “List of Title Word Abbreviations”
% http://www.issn.org/services/online-services/access-to-the-ltwa/) or
% the full name of the journal. 
% Citations and References in Supplementary files are permitted
% provided that they also appear in the reference list here. 

\externalbibliography{yes}
\bibliography{refs}

%%%%%%%%%%%%%%%%%%%%%%%%%%%%%%%%%%%%%%%%%%
\end{document}